\documentclass{aa}
\usepackage{graphicx}
\usepackage{deluxetable}
\usepackage{url}
\usepackage{natbib}
\usepackage{rotating}
\usepackage{amsmath}
\usepackage{txfonts}
\begin{document}

   \title{TANAMI: Tracking Active Galactic Nuclei with Austral Milliarcsecond Interferometry}

   \subtitle{I. First-Epoch 8.4\,GHz Images}

   \author{Roopesh Ojha
          \inst{1,2}
          \and
          Matthias Kadler\inst{3,4,5,6}
          \and
          Moritz B\"ock\inst{3,4}
          \and
          Roy Booth\inst{7}
          \and
          M. S. Dutka\inst{8}
          \and
          P. G. Edwards\inst{9}
          \and
          A. L. Fey\inst{1}
          \and
          L. Fuhrmann\inst{10}
          \and
          R. A. Gaume\inst{1}
          \and
          H. Hase\inst{11}
          \and
          S. Horiuchi\inst{12}
          \and
          D. L. Jauncey\inst{9}
          \and
          K. J. Johnston\inst{1}
          \and
          U. Katz\inst{4}
          \and
          M. Lister\inst{13}
          \and          
          J. E. J. Lovell\inst{14}
          \and
          C. M\"uller\inst{3,4}
          \and
          C. Pl\"otz\inst{15}
          \and
          J. F. H. Quick\inst{7}
          \and
          E. Ros\inst{16,10}
          \and
          G. B. Taylor\inst{17}
          \and
          D. J. Thompson\inst{18}
          \and
          S. J. Tingay\inst{19}
          \and
          G. Tosti\inst{20,21}
          \and
          A. K. Tzioumis\inst{9}          
          \and
          J. Wilms\inst{3,4}
          \and
          J. A. Zensus\inst{10}
          }

   \institute{United States Naval Observatory, 3450 Massachusetts Ave., NW,
             Washington DC 20392, U.S.A.\\
              \email{rojha@usno.navy.mil}
         \and
             NVI, Inc., 7257D Hanover Parkway, Greenbelt, MD 20770, USA
             \and
             Dr. Remeis Sternwarte, Astron. Institut der Universit\"at Erlangen-N\"urnberg, 
             Sternwartstrasse 7, 96049 Bamberg, Germany\\
          \email{matthias.kadler@sternwarte.uni-erlangen.de}
         \and 
             Erlangen Centre for Astroparticle Physics, Erwin-Rommel Str. 1, 91058 Erlangen, 
             Germany
         \and
             CRESST/NASA Goddard Space Flight Center, Greenbelt, MD 20771, USA
         \and 
             Universities Space Research Association, 10211 Wincopin Circle, Suite 500 Columbia, MD 
             21044, USA
           \and 
             Hartebeesthoek Radio Astronomy Observatory, PO Box 443, Krugersdorp 1740, South Africa
             \and
             The Catholic University of America, 620 Michigan Ave., N.E.,  Washington, DC 20064
             \and
             Australia Telescope National Facility, CSIRO, PO Box 76, Epping, NSW 1710, Australia
             \and
             Max-Planck-Institut f\"ur Radioastronomie, Auf dem H\"ugel 69, 53121 Bonn, Germany
             \and
             Bundesamt fur Kartographie und Geodesie, Germany entrusted with Transportable Integrated    	Geodetic Observatory, Universidad de Concepcion, Casilla 4036, Correo 3, Concepcion, Chile
             \and
             Canberra Deep Space Communication Complex, PO Box 1035, Tuggeranong, ACT 2901,  Australia
             \and
             Department of Physics, Purdue University, 525 Northwestern Avenue, West Lafayette, IN 47907, USA
             \and
             School of Mathematics \& Physics, Private Bag 37, University of Tasmania, Hobart TAS 7001, Australia
             \and
             Federal Agency for Cartography and Geodesy (BKG), Geodetic Observatory Wettzell, Sackenrieder Str. 25, 93444 Bad K\"otzting, Germany
             \and
             Dept. d'Astronomia i Astrof\'{\i}sica, Universitat de Val\`encia, E-46100 Burjassot, Val\`encia, Spain
             \and
             Department of Physics and Astronomy, University of New Mexico, Albuquerque NM, 87131, USA; Greg Taylor is also an Adjunct Astronomer at the National Radio Astronomy Observatory.
             \and
             Astrophysics Science Division, NASA Goddard Space Flight Center, Greenbelt, MD 20771, USA
             \and
             Curtin Institute of Radio Astronomy, Curtin University of Technology, Bentley, WA, 6102, Australia
             \and
             Istituto Nazionale di Fisica Nucleare, Sezione di Perugia, I-06123 Perugia, Italy
             \and
             Dipartimento di Fisica, Universit\`a degli Studi di Perugia, I-06123 Perugia, Italy
             }


 
  \abstract
   {A number of theoretical models vie to explain the $\gamma$-ray
   emission from active galactic nuclei (AGN). This was a key discovery
   of EGRET. With its broader energy coverage, higher resolution, wider
   field of view and greater sensitivity, the Large Area Telescope (LAT)
   of the \textsl{Fermi} Gamma-ray Space Telescope is dramatically
   increasing our knowledge of AGN $\gamma$-ray emission. However,
   discriminating between competing theoretical models requires
   quasi-simultaneous observations across the electromagnetic spectrum.
   By resolving the powerful parsec-scale relativistic outflows in
   extragalactic jets and thereby allowing us to measure critical
   physical properties, Very Long Baseline Interferometry observations
   are crucial to understanding the physics of extragalactic $\gamma$-ray
   objects.}
   {We introduce the TANAMI program (Tracking Active Galactic Nuclei with
   Austral Milliarcsecond Interferometry) which is monitoring an initial
   sample of 43 extragalactic jets located south of $-30$ degrees 
   declination at 8.4\,GHz and 22\,GHz since 2007. All aspects of the
   program are discussed. First epoch results at 8.4\,GHz are presented
   along with physical parameters derived therefrom.}
   {These observations were made during 2007/2008 using the telescopes of
   the Australian Long Baseline Array in conjunction with Hartebeesthoek
   in South Africa. These data were correlated at the Swinburne
   University correlator.}
   {We present first epoch images for 43 sources, some observed for the
   first time at milliarcsecond resolution. Parameters of these images as
   well as physical parameters derived from them are also presented and
   discussed. These and subsequent images from the TANAMI survey are
   available at \url{http://pulsar.sternwarte.uni-erlangen.de/tanami/} }
   {We obtain reliable, high dynamic range images of the southern
   hemisphere AGN.  All the quasars and BL Lac objects in the sample have
   a single-sided radio morphology. Galaxies are either double-sided,
   single-sided or irregular. About $28\%$ of the TANAMI sample has been
   detected by LAT during its first three months of operations. Initial
   analysis suggests that when galaxies are excluded, sources detected by
   LAT have larger opening angles than those not detected by LAT.
   Brightness temperatures of LAT detections and non-detections seem to
   have similar distributions. The redshift distributions of the TANAMI
   sample and sub-samples are similar to those seen for the bright
   $\gamma$-ray AGN seen by LAT and EGRET but none of the sources with a
   redshift above 1.8 have been detected by LAT.}


  \authorrunning{Ojha et al.}
  \titlerunning{TANAMI First Epoch 8.4\,GHz Images}
  \maketitle
%

\section{Introduction\label{sec:intro}}

Blazars are a radio-loud, violently variable, high-luminosity,
high-polarization subset of Active Galactic Nuclei (AGN) that show
luminosity variations across the electromagnetic spectrum. They typically
exhibit apparent superluminal motion along the innermost parsecs of their
radio jets. Their observed behaviour suggests they are very compact
objects with parsec-scale jets oriented close to our line of sight.
Despite decades of observations and modeling of the powerful relativistic
jets of AGN, many fundamental questions about them remain unanswered. Jet
composition, formation, and collimation are not well understood. {\bf
Neither} are the mechanisms responsible for their propagation, radiation,
and interaction with their ambient medium. 

The discovery by the EGRET detector onboard the Compton Gamma Ray
Observatory (CGRO) that blazars can be bright $\gamma$-ray emitters
\citep{Hartman1992} was a major breakthrough in the study of AGN. Study
of the $\gamma$-ray blazar population suggested that there is a close
connection between $\gamma$-ray and radio emission, spectral changes,
outbursts and the ejection of parsec-scale jet components
\citep{Dondi1995}. Many, but not all, of the radio brightest blazars have
been detected at $\gamma$-ray energies. On the other hand, a small number
of extragalactic jets which do not belong to the blazar class have been
shown to be bright $\gamma$-ray sources, as well \citep[e.g., NGC\,1275
and Cen A;][]{AbdoLBAS2009}. Currently there are a number of models
attempting to explain the observed $\gamma$-ray emission of blazars and
other extragalactic jets \citep[see][for a review]{Bottcher2007}.
Simultaneous broadband spectral energy distribution (SED) measurements
across the electromagnetic spectrum are required to discriminate between
such models.  

Very Long Baseline Interferometry (VLBI) monitoring of blazars is a
crucial component of this multiwavelength suite of observations as it
provides the only direct measure of relativistic motion in AGN jets
allowing us to calculate intrinsic jet parameters such as jet speed,
Doppler factor, opening and inclination angles. They also provide the
possibility of identifying the location and extent of emission regions.
Thus VLBI observations constrain numerical jet simulations and provide
tests of the relativistic-beam model \citep[e.g.,][]{Cohen2007} that are
not possible with any other observational technique. 

The successful launch of the \textsl{Fermi} Gamma-ray Space Telescope,
formerly known as GLAST \citep[Gamma-ray Large Area Space
Telescope;][]{Gehrels1999}, on June 11th, 2008 was a major milestone in
the quest to understand the connection between the low- and high-energy
sections of blazar SEDs. The Large Area Telescope
\citep[LAT;][]{Atwood2009} instrument on \textsl{Fermi} has broader
energy coverage (20\,MeV -- 300\,GeV), higher resolution, wider field of
view (over $20\%$ of the sky) and a sensitivity $\sim$30 times greater
than EGRET. The combination of its wide field of view with a scanning
pattern of observation means LAT observes the entire sky every $\sim$3
hours making LAT capable of monitoring the sky on timescales from hours
to years. Its higher sensitivity at higher energies allows LAT to measure
high energy cutoffs which elucidate acceleration mechanisms, radiation
and magnetic fields at the source. 


The \textbf{T}racking \textbf{A}ctive Galactic \textbf{N}uclei with
\textbf{A}ustral \textbf{M}illiarcsecond \textbf{I}nterferometry (TANAMI)
program is a parsec-scale radio monitoring program targeting
extragalactic jets south of $-$30 degrees declination. It uses the
telescopes of the Australian Long Baseline Array \citep[LBA,
e.g.,][]{Ojha2004b}, and other telescopes in South Africa, Antarctica and
Chile to monitor an initial sample of 43 sources at approximately 2-month
intervals. The observations are typically made at two frequencies,
8.4\,GHz and 22\,GHz, in order to calculate the spectral indices for the
core as well as bright jet components. TANAMI began observations in
November 2007 before the launch of \textsl{Fermi} so a proper observation
cadence for the targets could be determined. To maximize the usage of
TANAMI data, images of sources are made available at
\url{http://pulsar.sternwarte.uni-erlangen.de/tanami/} as soon as they
are available. The current target list of TANAMI is maintained at
\url{http://pulsar.sternwarte.uni-erlangen.de/tanami/sample/}.

While the immediate driver for TANAMI was the imminent launch of the
\textsl{Fermi} Gamma-ray Space Telescope, these dual-frequency
observations of morphology, motion, and other temporal variations of the
most poorly studied third of the AGN sky have a number of other
applications. One particularly exciting prospect is the possibility of
combining $\gamma$-ray and radio monitoring of parsec-scale jet
kinematics (and production) with a sensitive neutrino telescope.  The
ANTARES neutrino telescope (see \url{http://antares.in2p3.fr}), which is
now fully operational and the KM3NeT detector (which has a projected
completion date of 2011, see \url{http://www.km3net.org}) are being
designed to detect neutrino point sources\footnote{See KM3NeT Conceptual
Design Report and references therein at
\url{http://www.km3net.org/CDR/CDR-KM3NeT.pdf}}.  Extragalactic jets are
among the most promising candidates to be neutrino point sources
\citep{Waxman2007}. Combined TANAMI, \textsl{Fermi} and ANTARES data will
be used to search for neutrino signals correlated with $\gamma$-ray
flares and epochs of jet production. 
The results of this work will be the subject of future publications.

In this paper, we describe how we selected our initial source sample
(Sect.~\ref{sec:sample}) and explain our observation and data reduction
procedures (Sect.~\ref{sec:observations}).  We then present the first
epoch images of 43 target sources at 8.4\,GHz in Sect.~\ref{sec:results}
followed by brief notes on individual sources
(Sect.~\ref{sec:notesindiv}). This is followed by a discussion of our
results (Sect.~\ref{sec:discussion}) and we end with our conclusions
(Sect.~\ref{sec:conclusions}). Throughout the paper we use the
cosmology $H_0$=73\,km\,s$^{-1}$\,Mpc$^{-1}$, $\Omega_{m}$=0.27,
$\Omega_{\Lambda}$=0.73 where the symbols have their traditional
meanings. 

\section{Definition of the Sample \label{sec:sample}} The TANAMI sample
has been defined as a hybrid radio and $\gamma$-ray selected sample of
AGN south of $\delta=-30^\circ$. Its main components are {\sc i)} a radio
selected flux-density limited subsample and {\sc ii)} a $\gamma$-ray
selected subsample of known and candidate $\gamma$-ray sources based on
results of \textsl{CGRO}/EGRET. The radio subsample includes all sources
(within our declination range) from the catalogue of \citet{Stickel1994}
above a limiting radio flux density of S$_{\rm 5\,GHz}>2$\,Jy, which have
a flat radio spectrum ($\alpha>-0.5$, $\mathrm{S}\sim\nu^{+\alpha}$) between
2.7\,GHz and 5\,GHz. The 21 sources selected according to this radio
flux-density criterion represent a Southern-hemisphere extension of the
MOJAVE\,1 sample \citep{Lister2009}, which is complete in the declination
range $0^\circ>\delta>-20^\circ$ down to the same flux-density limit at
15\,GHz. The $\gamma$-ray selected subsample includes all known
$\gamma$-ray blazars detected by EGRET south of $\delta=-30^\circ$, both
the high-confidence and low-confidence associations made by
\citet{Hartman1999}, \citet{Tornikoski2002}, \citet{Sowards-Emmerd2004}
and \citet{Bignall2008}. In addition, we have also included 4 known
intra-day variable (IDV) sources (0405$-$385, 1144$-$379, 1257$-$326, and
1323$-$526) and 8 other sources, which either share the radio properties
of EGRET-detected blazars at lower radio flux density or which represent
prototypical examples of other AGN classes such as the bright and nearby
radio galaxy Pictor\,A (0518$-$458) or the gigahertz peaked spectrum
(GPS) source NGC\,6328 (1718$-$649). In total, this sample contains 44
objects\footnote{The candidate EGRET $\gamma$-ray blazar 0527$-$359 was
not observed due to scheduling problems.}. Many of the sources in this
sample have been well studied with VLBI in the past \citep{Tingay1996b,
Shen1997, Shen1998, Tingay2002, Ojha2004, Ojha2005, Scott2004,
Horiuchi2004, Dodson2008} but for about $30\,\%$ only very limited
information at typically much lower resolutions and image fidelity are
available in the literature.

Most AGN with bright compact radio emission are strongly variable. This
led the MOJAVE team to define their statistically complete radio-selected
sample based on radio light curve observations of a large sample of
sources over a ten-year time baseline and to include all sources that
exceeded their flux-density limit at any epoch during this period. In
contrast to this, the TANAMI sample is not statistically complete.
However, based on experience from the transition of the original VLBA
2\,cm Survey sample into the MOJAVE sample \citep{Lister2009}, we can
consider the radio-selected subsample of TANAMI as being representative
of a complete sample.  In a similar sense, the \textsl{Fermi} LBAS (LAT
bright AGN sample) list is not statistically complete 
\citep{AbdoLBAS2009} and neither is the subsample of LBAS-TANAMI sources.
A few bright $\gamma$-ray sources have no or only low-confidence
associations\footnote{\citet{AbdoBRIGHT2009} list three sources below 
declination $-30$\,degrees and at high galactic latitude
$|b|>10$\,degrees, which have no association whatsoever and might be
unknown AGNs: 0FGL~J\,0614.3$-$3330, 0FGL~J\,1311.9$-$3419, and
0FGL~J\,2241.7$-$5239. Two sources in this region do have only
low-confidence associations with AGN (0FGL~J\,0407.6$-$3829 associated
with PKS~0405$-$385 and 0FGL~J\,0412.9$-$5341 associated with
PMN~J\,0413$-$5332).} and in addition, the presence of known EGRET
sources, which are not in the LBAS\footnote{Below declination
$-30$\,degrees, there are three 3EG sources \citep{Hartman1999}, which
are not part of the LBAS sample: 0454$-$463, 1424$-$418, and 1933$-$400.
nine other sources associated with lower confidence with EGRET sources by
\citet{Tornikoski2002} and \citet{Sowards-Emmerd2004} have also not been
seen as bright $\gamma$-ray sources by LAT during its first three
months.}, indicates the effect of long-term variability being also
important in the $\gamma$-ray regime. We are working on improving the
completeness of the $\gamma$-ray selected TANAMI subsample in
collaboration with the \textsl{Fermi}/LAT science team. In addition to
the 44 sources of the initial TANAMI sample, we have begun observations
of 19 additional \textsl{Fermi} sources in November 2008, which will be
reported elsewhere. Most of these sources have not been observed with
VLBI before. We plan to continue adding new \textsl{Fermi}-detected
sources to the TANAMI list through 2009.

The Veron-Veron 12th edition catalog \citep{VeronVeron2006} was used to
obtain optical classifications. The sample contains 24 quasars (optically
unresolved broad-emission line objects), 6 BL\,Lac objects (optically
unresolved sources with weak or absent emission lines), and 10 radio
galaxies (optical galaxies, which are identified with radio sources).
Three sources are unclassified. 

After the beginning of \textsl{Fermi} science operations and the
publication of the first list of bright $\gamma$-ray emitting AGN
detected by \textsl{Fermi}/LAT between August and October 2008
\citep{AbdoLBAS2009}, our sample-selection criteria have proved to be
highly efficient in picking up bright $\gamma$-ray emitters. Our original
sample contains 10 of the 18 high-confidence associations of the LBAS
sample within our declination range, one out of two lower confidence
association and one additional bright $\gamma$-ray source association
(1759$-$396) outside the galactic latitude range considered by
\citet{AbdoBRIGHT2009} . Four of these 12 sources have entered the TANAMI
sample because they belong to both the radio flux-density limited sample
and to the EGRET sample, two and four belong only to the radio or EGRET
sample, respectively, and two belong to the IDV class. Details are given
in Table~\ref{table:sourcelist}. Notably, one out of only two LBAS radio
galaxies belongs to the TANAMI sample (Cen\,A), as well as two out of
seven high-frequency peaked BL\,Lac objects, which are also detected in
the TeV energy range (2005$-$489 and 2155$-$304).

\section{Observations and Data Reduction\label{sec:observations}}

TANAMI observations are made using the five telescopes of the Australian
Long Baseline Array (LBA\footnote{The Long Baseline Array is part of the
Australia Telescope which is funded by the Commonwealth of Australia for
operation as a National Facility managed by CSIRO.}) along with other
affiliated telescopes. Within Australia, TANAMI has periodic access to
the 70\,m and the 34\,m telescopes at NASA's Deep Space Network (DSN)
located at Tidbinbilla, near Canberra in the Australian Capital Territory
(ACT). When available, these telescopes add crucial sensitivity to map
details of the jet structure with higher fidelity, besides improving the
$(u,v)$-coverage. Through September 2008, the highest resolution,
intercontinental baselines were provided by the 26\,m telescope in
Hartbeesthoek, South Africa. However, this telescope experienced a major
failure of a polar shaft bearing in October 2008 and is likely to remain
unavailable for some time (Jonathan Quick, personal communication).
Fortunately, through a successful International VLBI Service (IVS)
proposal we have obtained access to two telescopes, O'Higgins, Antarctica
and TIGO (Transportable Integrated Geodetic Observatory), Chile. Both
these telescopes are operated by the Bundesamt f\"{u}r Kartographie und
Geod\"{a}sie (BKG), the federal agency responsible for cartography and
geodesy in Germany. These two telescopes greatly improve our
$(u,v)$-coverage and partially offset the loss of Hartebeesthoek. Details
of all of these telescopes used in TANAMI observations are summarized in
Table~\ref{table:antennas}. The telescopes participating in the
observations reported in this paper are indicated in
Table~\ref{table:epochs}.


TANAMI observations are made at two frequencies: 8.4\,GHz (X-band) and at
22\,GHz (K-band). Imaging observations at 8.4\,GHz generally yield the
best image fidelity and thus the most detailed structural information
with this array as demonstrated, for example, by the U.S. Naval
Observatory/ATNF, International Celestial Reference Frame (ICRF) imaging
program \citep{Ojha2004,Ojha2005}. This frequency is high enough to
provide good resolution and low enough to avoid missing extended
structure that typically has a steep spectrum. Atmospheric effects are
also negligible at this frequency.  Imaging observations at 22\,GHz are
challenging but feasible \citep{Tingay2003}. Not only do 22\,GHz images
show morphology closer to the cores of the AGN, in combination with the
8\,GHz images, they yield spectral information on the core and bright jet
components. These are a critical component of the broadband SEDs needed
to understand the energetics of AGN. One 24-hour epoch at each frequency
is observed approximately every two months. In this paper, we describe
the 8.4\,GHz observations. Table~\ref{table:epochs} summarizes the
observations which are reported here. 

For the array described above, the $\it typical$ angular resolution
(synthesized beam) achieved by TANAMI is 1.5--4 by 0.5--1.0\,mas in size,
with the highest resolution in the east-west direction. Each source was
observed in about 6 scans of approximately 10 minutes each. The data were
recorded on the LBADRs (Long Baseline Array Disk Recorders) and
correlated on the DiFX software correlator \citep{Deller2007} at
Swinburne University in Melbourne, Victoria. From November 2008 on, the
data is being correlated at Curtin University in Perth, Western
Australia.

The correlated data were loaded into AIPS using the locally written task
ATLOD which is needed to read the data in the format that the LBA
generates. Thereafter, data inspection, initial editing and fringe
fitting was done in the standard manner using the National Radio
Astronomy Observatory's Astronomical Image Processing System (AIPS)
software \citep{Greisen1988}. Observations of known sources with $\ge
90\%$ of their correlated flux in a compact core were used to improve
overall amplitude calibration. For each antenna, a single amplitude gain
correction factor was derived based on fitting a simple Gaussian source
model to the visibility data of the respective compact source after
applying only the initial calibration based on the measured system
temperatures and gain curves. Based on the differences between the
observed and model visibilities, gain correction factors were calculated
and the resulting set of amplitude gain correction factors was then
applied to the visibility data of the target sources. The accuracy of the
absolute amplitude calibration is conservatively estimated to be $20\%$. 

The imaging was performed applying standard methods using the program 
{\sc difmap} \citep{Shepherd1997}. Specifically, the data were averaged
into 30 second bins and then imaged using the {\sc clean} algorithm,
giving the same weight to all visibility data points (natural weighting)
and making use of phase self-calibration.  The best model that could be
obtained in this initial cycle of the hybrid mapping process was used to
self calibrate the visibility amplitudes by applying time-independent
gain factors for each antenna in the array. The model was then cleared
and the resulting improved data were imaged in additional hybrid-mapping
cycles following the same strategy but using time-dependent gain factor
functions with subsequently smaller time intervals (180, 60, 20, 5, 2, 1
minutes). Before beginning a new cycle, the data were examined and edited
if necessary. The images shown in Fig.~\ref{fig:tan_page1} through
Fig.~\ref{fig:tan_page8} result from the final hybrid-mapping cycle
using natural weighting and a 30 second solution interval. Some sources
exhibit diffuse large-scale emission which could only be recovered by
down-weighting the data on the longest baselines ($(u,v)$-tapering). In
Fig.~\ref{fig:tan_taper_page1}, Fig.~\ref{fig:tan_taper_page2} and
Fig.~\ref{fig:tan_taper_page3} we show tapered images for 13 sources.

\section{Results \label{sec:results}}

Physical characteristics of the TANAMI sources, where available, are
summarized in Table~\ref{table:sourcelist}. The table lists their IAU
source designation followed by their alternate name (where appropriate)
and their Right Ascension and Declination in J2000.0 coordinates. Their
optical identification, $V$-magnitude and redshift are listed in
successive columns. The last four columns indicate which sources belong
to the radio and $\gamma$-ray selected subsamples and whether they were
detected by EGRET and LAT, respectively, where the LAT-detection flag
refers to the bright-source list of \citet{AbdoLBAS2009} based on the
first three months of \textsl{Fermi} observations. 

\begin{figure*}
   \centering
   \resizebox{0.92\hsize}{!}{\includegraphics{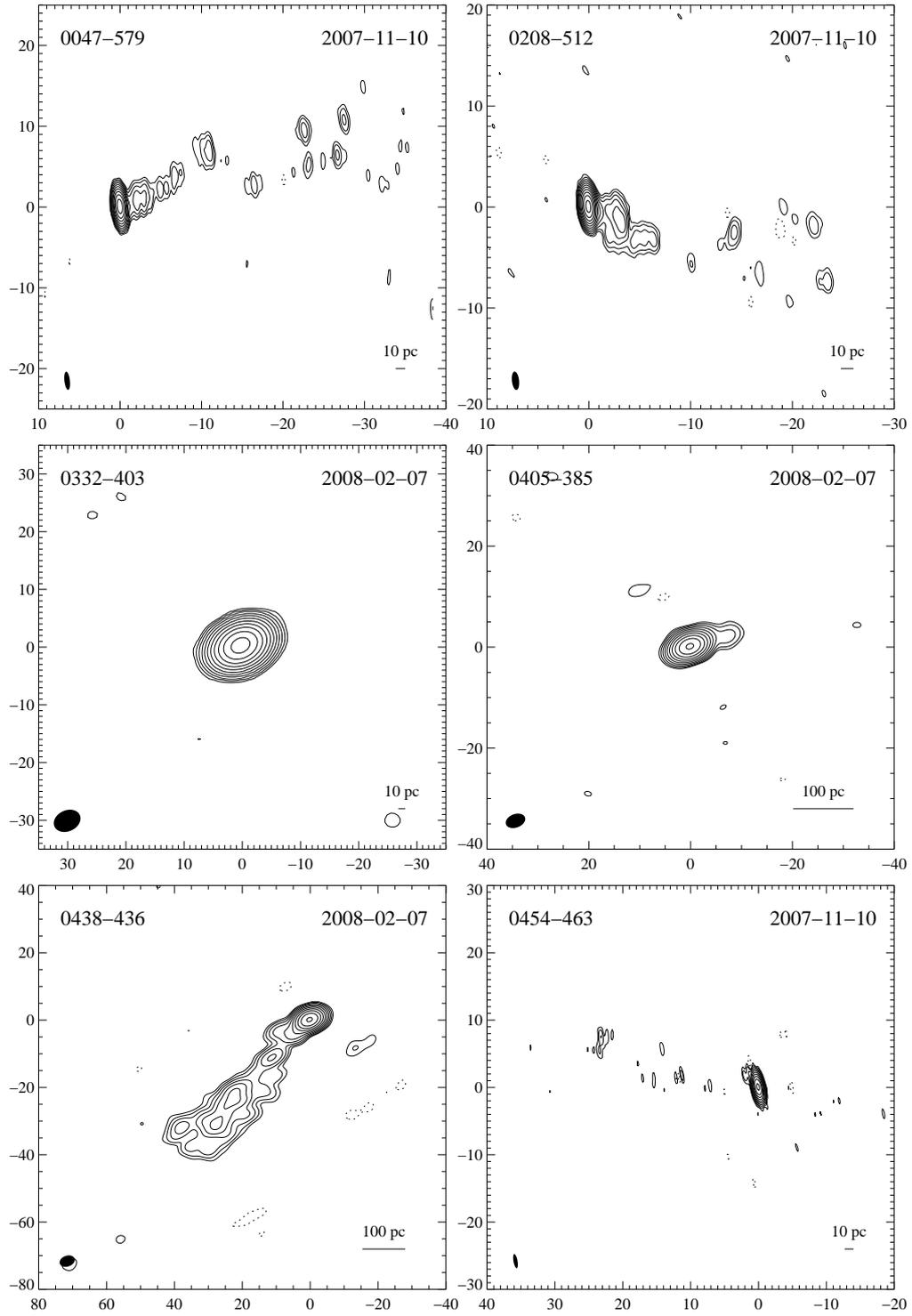}}
   \caption{Contour maps of the 43 TANAMI sources at 8.4\,GHz.  The
scale of each image is in milliarcseconds. The FWHM Gaussian restoring
beam applied to the images is shown as a hatched ellipse in the lower
left of each panel. Each panel also shows a bar representing a linear
scale between 0.1\,pc and 100\,pc except for the sources without 
a measured redshift. }
              \label{fig:tan_page1}
    \end{figure*}

\begin{figure*}
   \centering
   \resizebox{0.92\hsize}{!}{\includegraphics{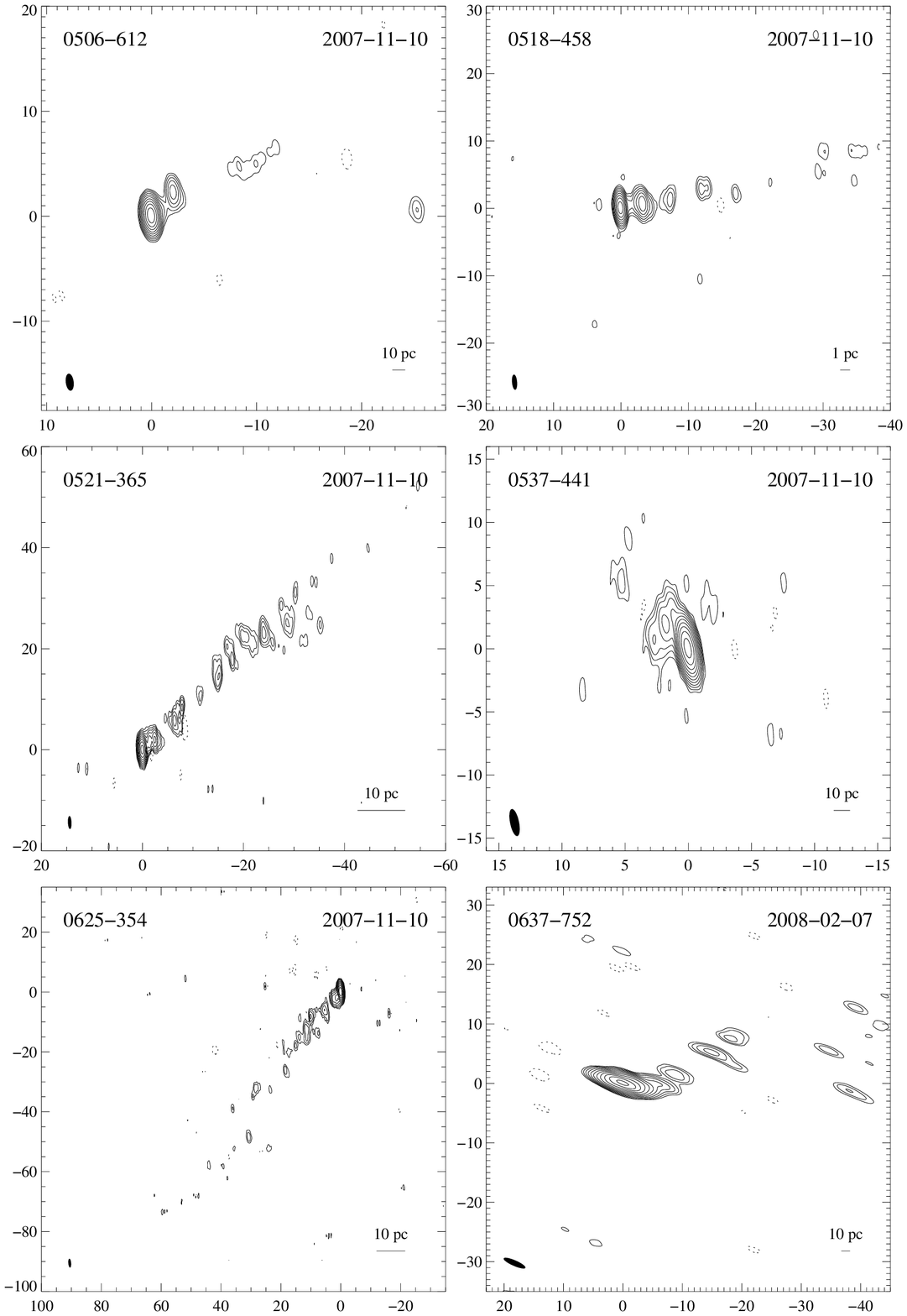}}
   \caption{Same as Fig. 1}
              \label{fig:tan_page2}
    \end{figure*}

\begin{figure*}
   \centering
   \resizebox{0.92\hsize}{!}{\includegraphics{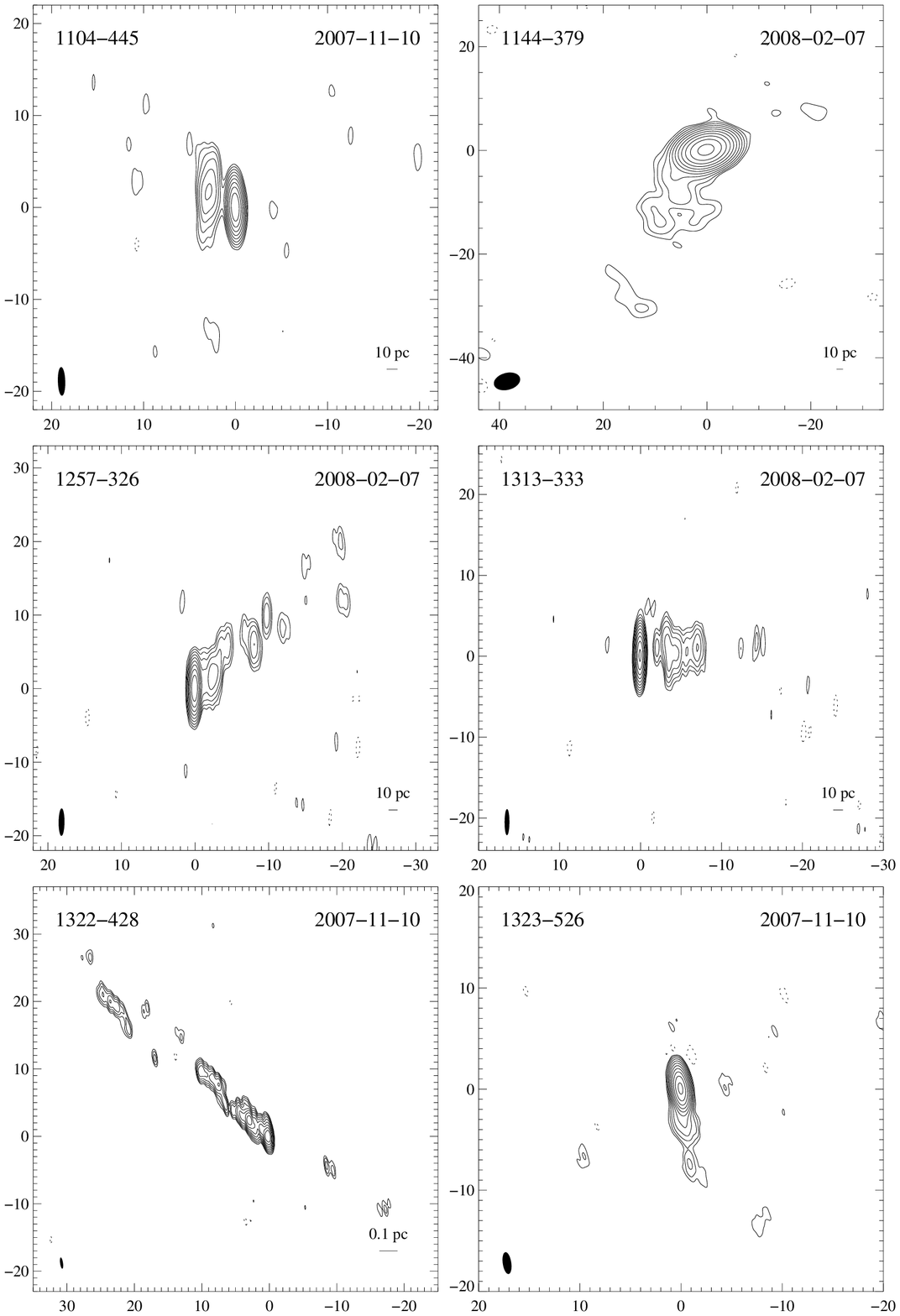}}
   \caption{Same as Fig. 1}
              \label{fig:tan_page3}
    \end{figure*}

\begin{figure*}
   \centering
   \resizebox{0.92\hsize}{!}{\includegraphics{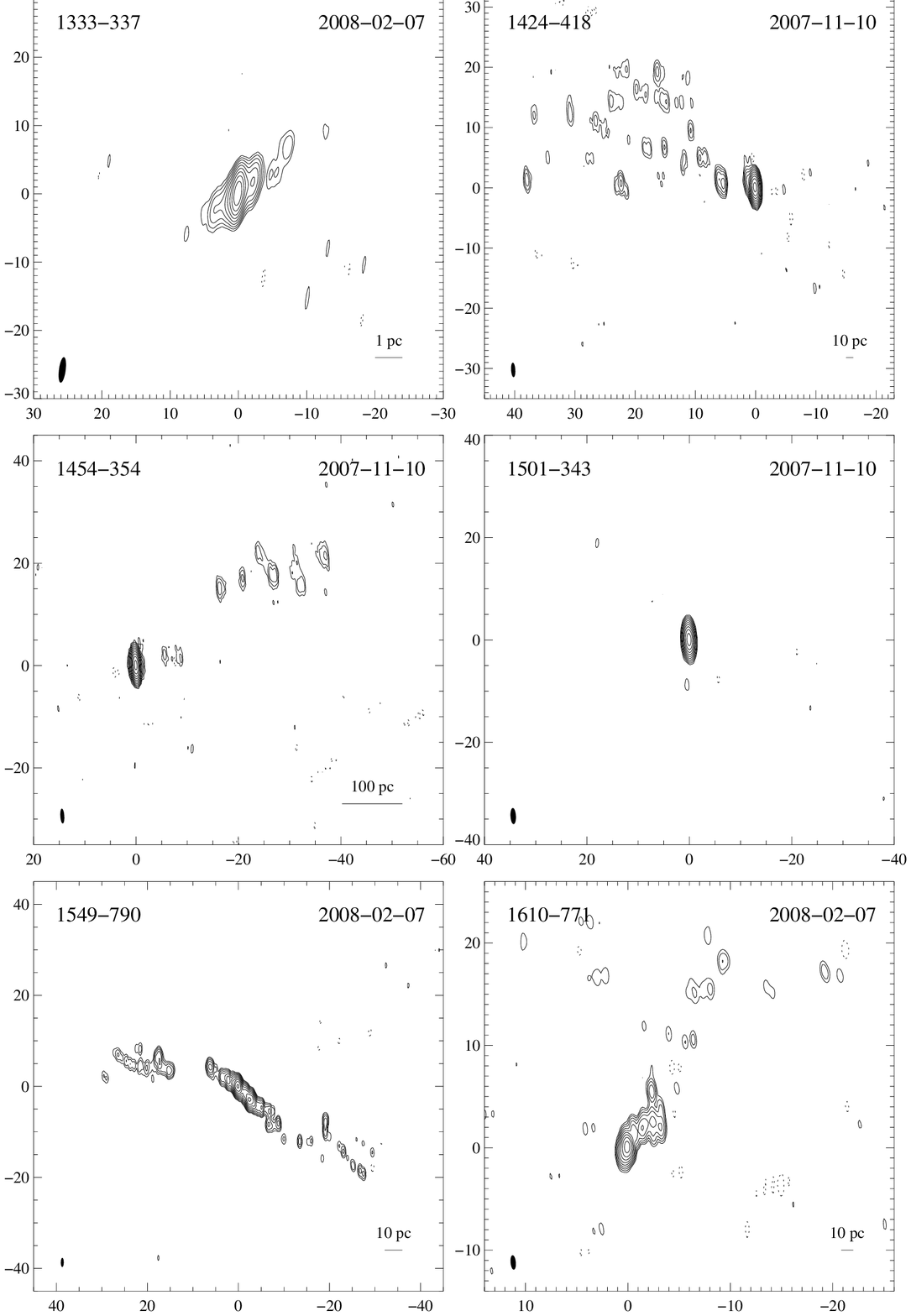}}
   \caption{Same as Fig. 1}
              \label{fig:tan_page4}
    \end{figure*}

\begin{figure*}
   \centering
   \resizebox{0.92\hsize}{!}{\includegraphics{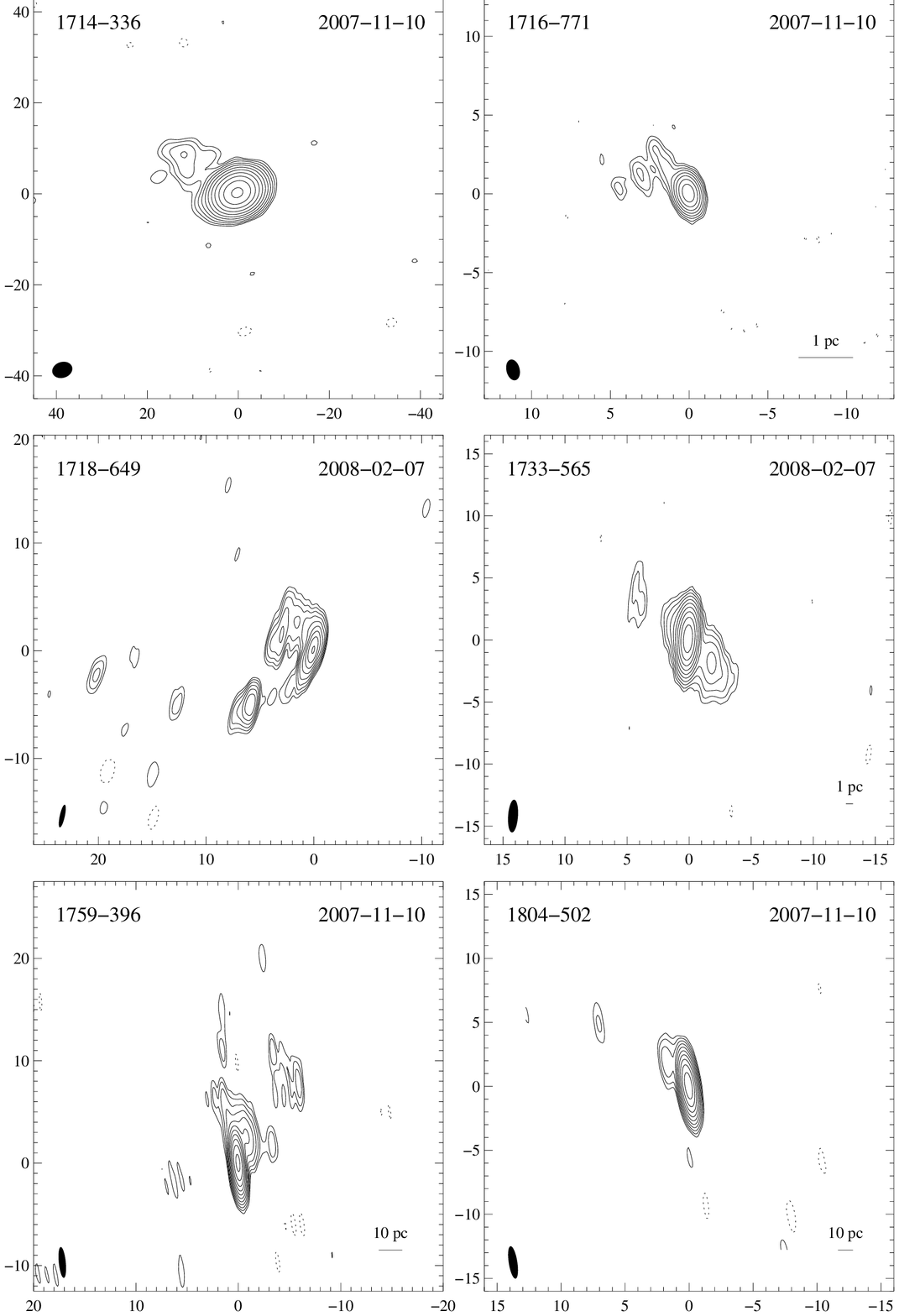}}
   \caption{Same as Fig. 1}
              \label{fig:tan_page5}
    \end{figure*}

\begin{figure*}
   \centering
   \resizebox{0.92\hsize}{!}{\includegraphics{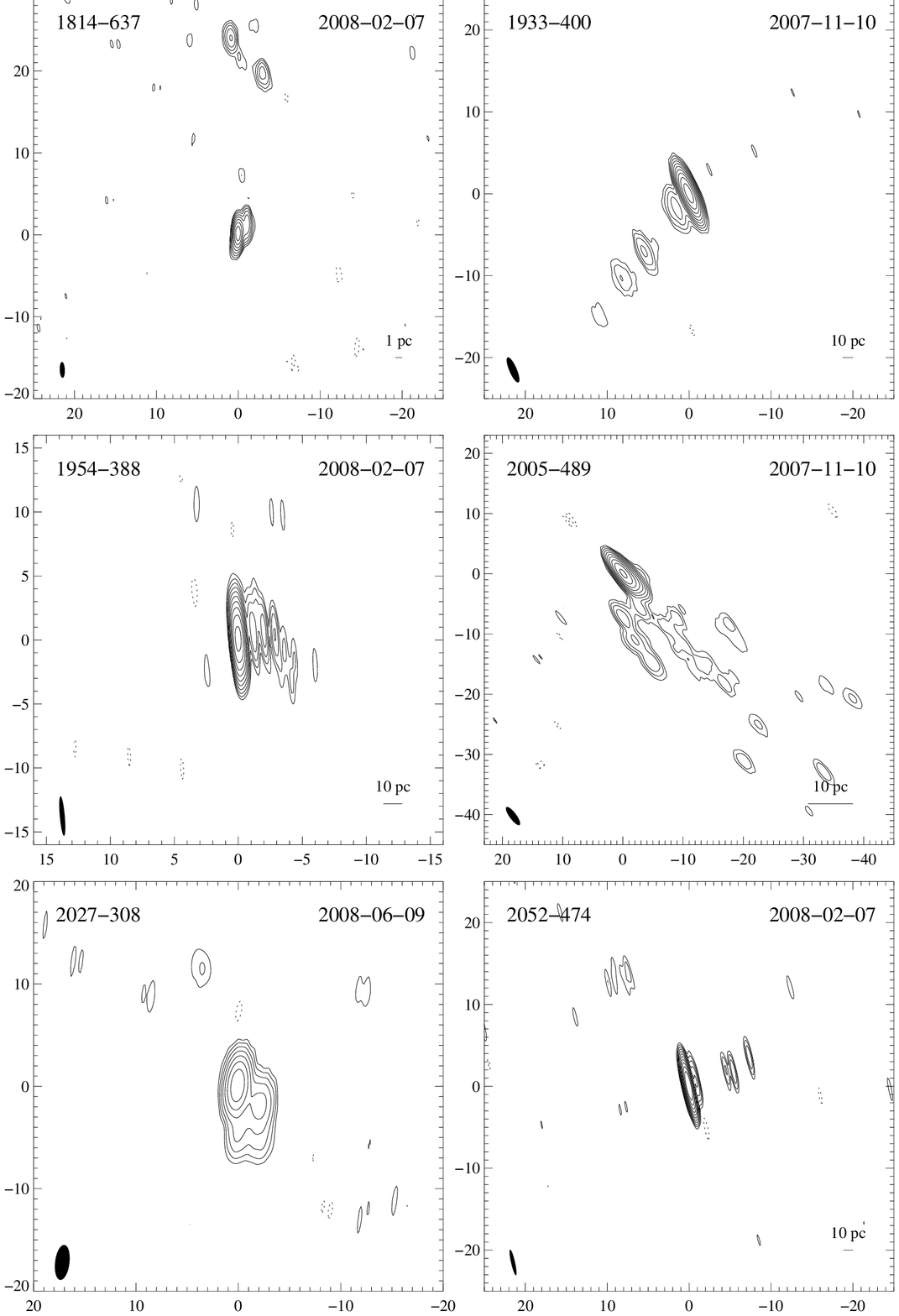}}
   \caption{Same as Fig. 1}
              \label{fig:tan_page6}
    \end{figure*}

\begin{figure*}
   \centering
   \resizebox{0.92\hsize}{!}{\includegraphics{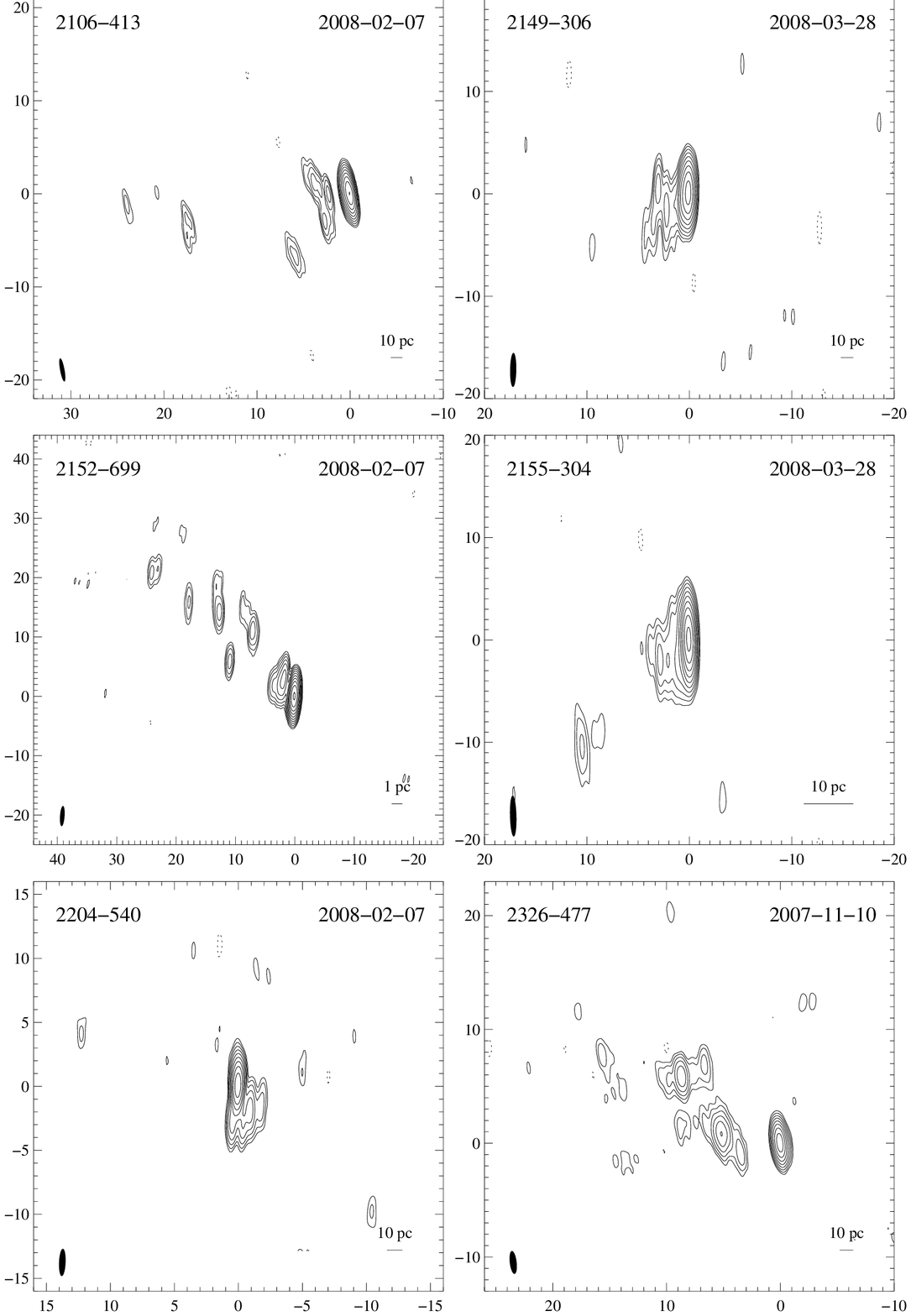}}
   \caption{Same as Fig. 1}
              \label{fig:tan_page7}
    \end{figure*}

\begin{figure*}
   \centering
   \resizebox{0.92\hsize}{!}{\includegraphics{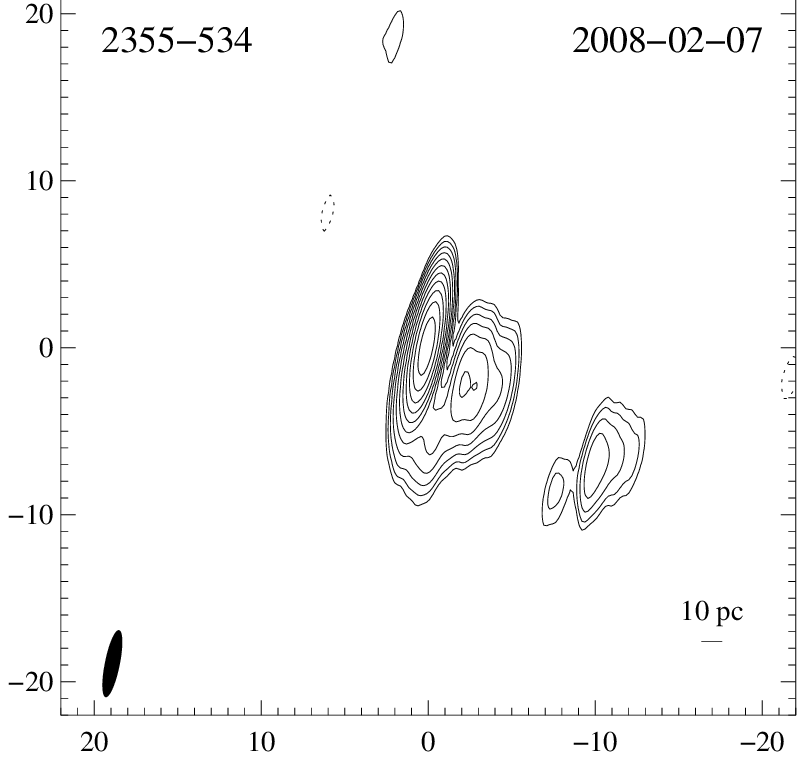}}
   \caption{Same as Fig. 1}
              \label{fig:tan_page8}
    \end{figure*}

\begin{figure*}
   \centering
   \resizebox{0.92\hsize}{!}{\includegraphics{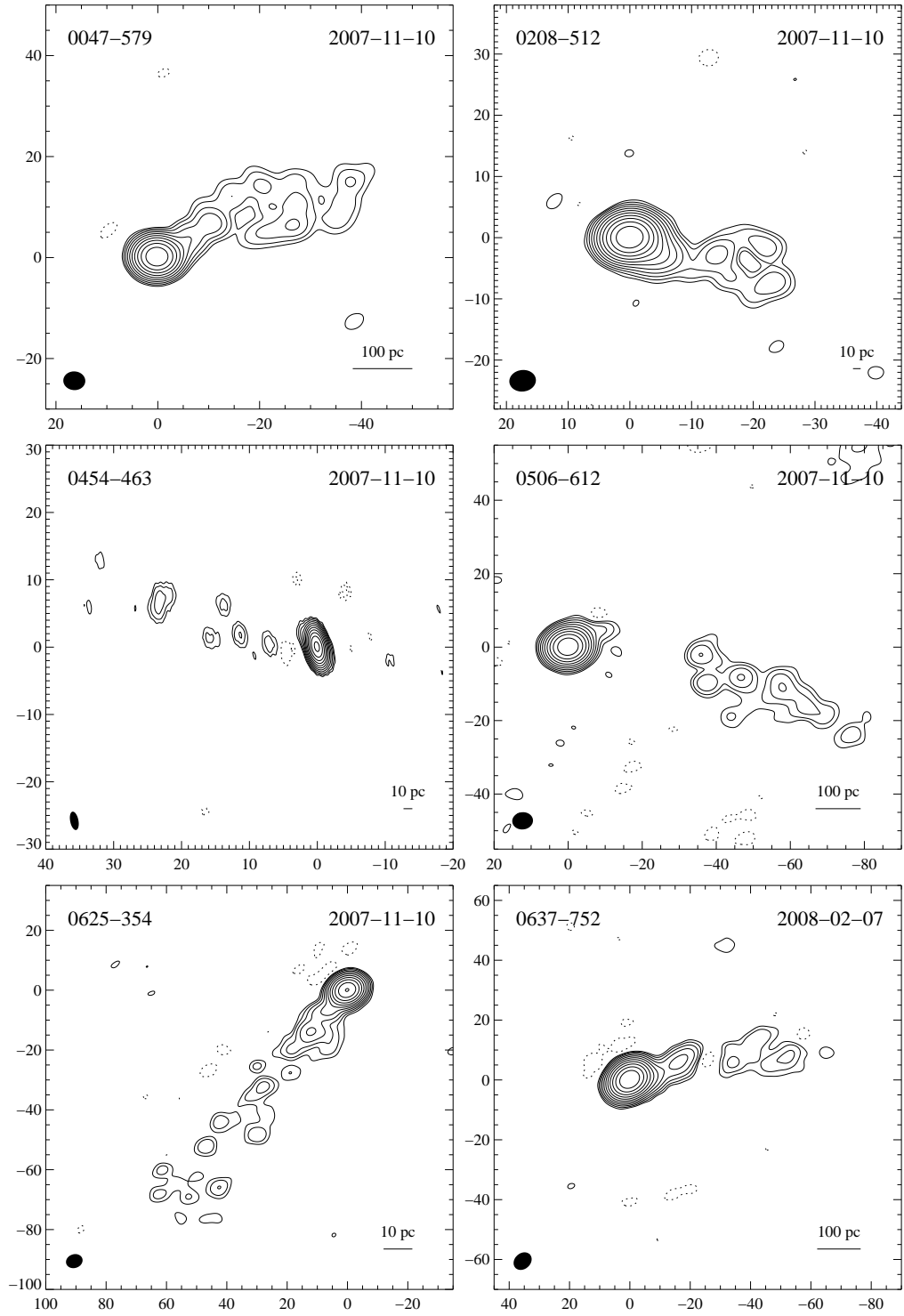}}
   \caption{Tapered contour maps of 13 TANAMI sources at 8.4\,GHz revealing more 
extended emission than visible in the naturally-weighted images.  The
scale of each image is in milliarcseconds. The FWHM Gaussian restoring
beam applied to the images is shown as a hatched ellipse in the lower
left of each panel. Each panel also shows a bar representing a linear
scale of 1\,pc or 10\,pc except for the sources without 
a measured redshift. }
              \label{fig:tan_taper_page1}
    \end{figure*}

\begin{figure*}
   \centering
   \resizebox{0.92\hsize}{!}{\includegraphics{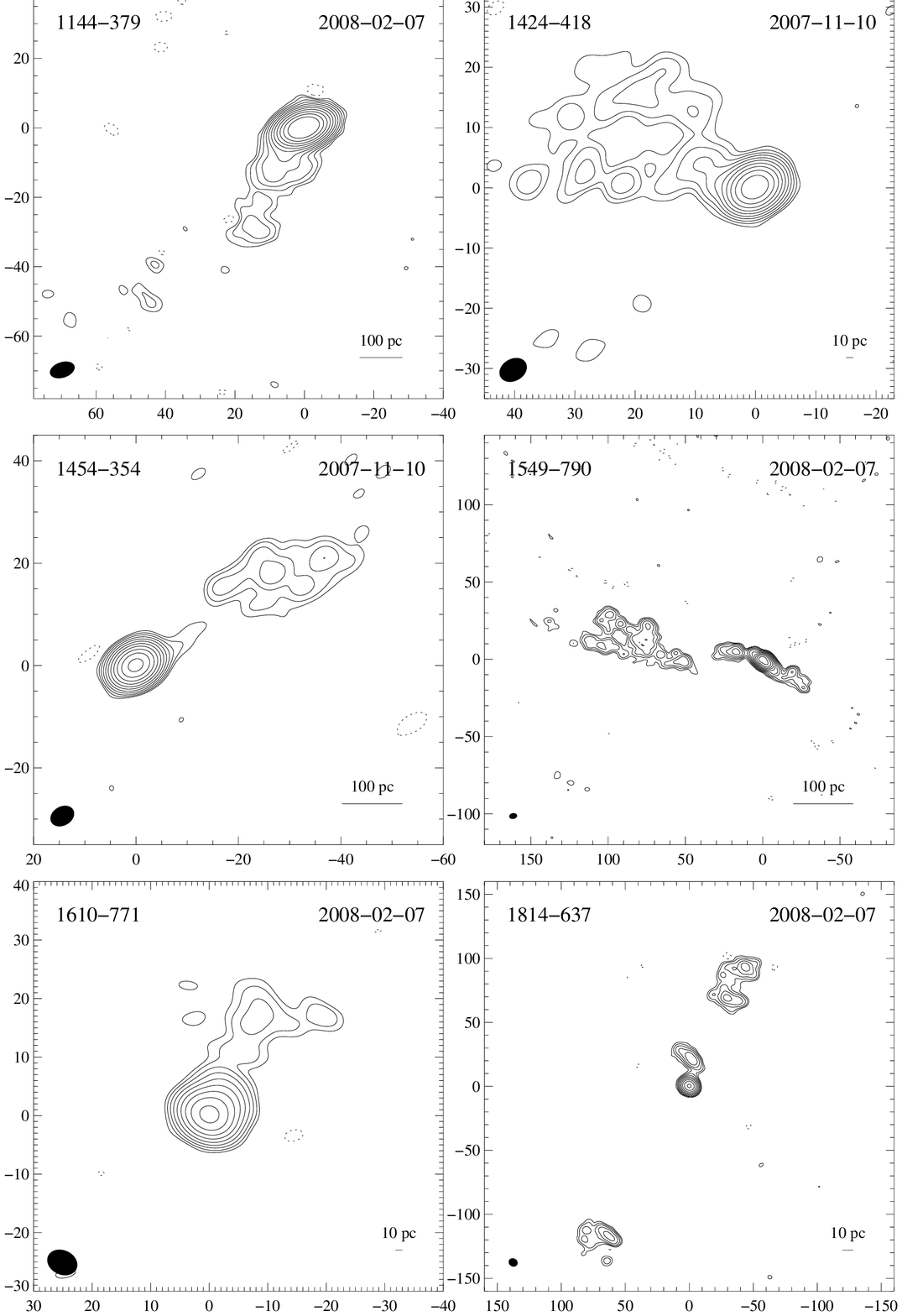}}
   \caption{Same as Fig. 9}
              \label{fig:tan_taper_page2}
    \end{figure*}

\begin{figure*}
   \centering
   \resizebox{0.92\hsize}{!}{\includegraphics{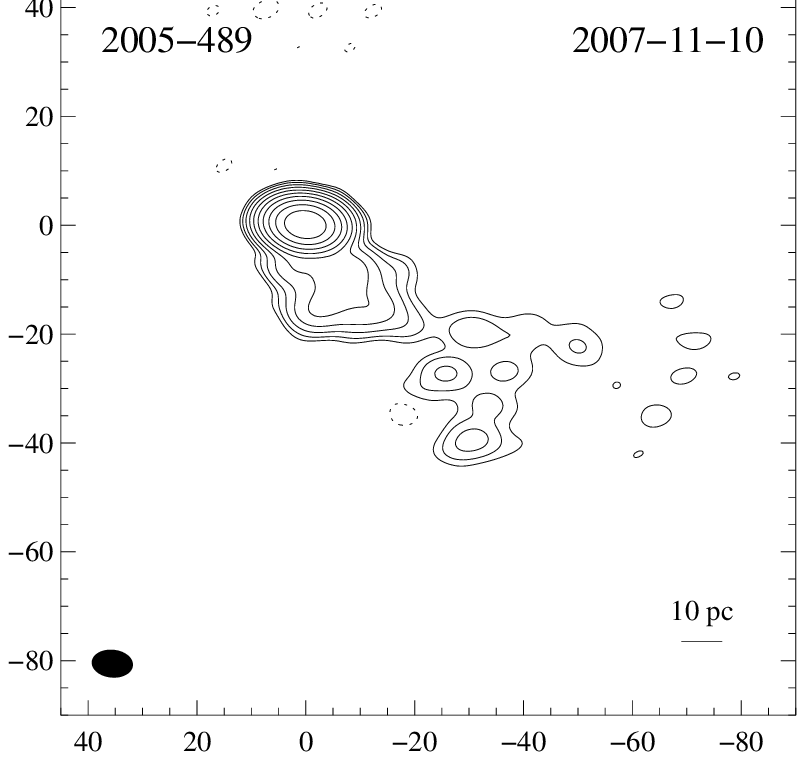}}
   \caption{Same as Fig. 9}
              \label{fig:tan_taper_page3}
    \end{figure*}

Figure~\ref{fig:tan_page1} through Fig.~\ref{fig:tan_page8} show contour
plots of the 43 sources of the initial TANAMI sample.  These images are
made with natural weighting. A subset of the sources that have diffuse
large-scale emission are shown in Fig.~\ref{fig:tan_taper_page1},
Fig.~\ref{fig:tan_taper_page2} and Fig.~\ref{fig:tan_taper_page3} using
$(u,v)$-tapering. The scale of each image is in milliarcseconds. The FWHM
Gaussian restoring beam applied to the images is shown as a hatched
ellipse in the lower left of each panel. Each panel also shows a bar
representing a linear scale of 1\,pc, 10\,pc, or 100\,pc depending on the
source extent and distance, except for the sources without a measured
redshift. The average root-mean-square (rms) noise in the images is
$\sim$0.43\,mJy beam$^{-1}$ with a median rms of $\sim$0.33\,mJy
beam$^{-1}$.

Image parameters are listed in Table~\ref{table:sourcestructure}.
The first two columns list the IAU source name and the epoch of the image
shown.  The lowest contour level is at 3 times the root-mean-square noise
and is listed in Col.\,(3). The peak flux density in each image and the
total flux are given in Col.\,(4) and Col.\,(5). The major axis, minor
axis, and position angle of the restoring beam are in Cols.\,(6)--(8).
Column\,(9) and Col.\,(10) show the structural classification and the
core brightness temperature respectively.  Both of these were evaluated
using criteria described in Sect.~\ref{sec:discussion}.  The final two
columns describe the core luminosity and total luminosity of each source.
Table~\ref{table:sourcestructuretapered} summarizes the image parameters
for the 13 sources for which we present tapered images. The first eight
columns are identical to those of Table~\ref{table:sourcestructure}. The
last column indicates the baseline length in M$\lambda$ at which the
visibility data were down weighted to $10\%$. 

\begin{figure*}
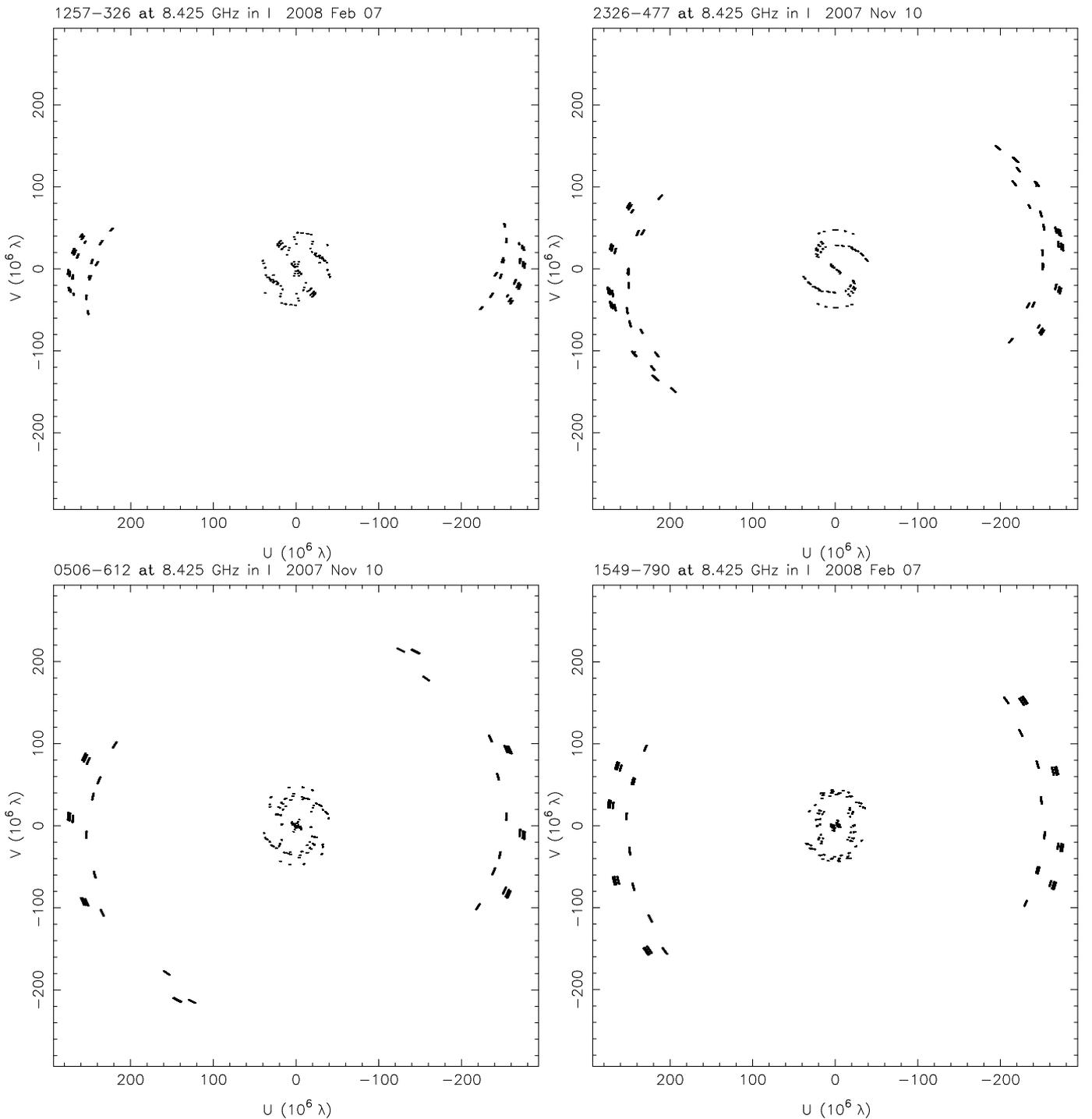

\centering
\resizebox{0.49\hsize}{!}{\includegraphics[angle=-90]{12724fg12a.ps}}
\resizebox{0.49\hsize}{!}{\includegraphics[angle=-90]{12724fg12b.ps}}
\resizebox{0.49\hsize}{!}{\includegraphics[angle=-90]{12724fg12c.ps}}
\resizebox{0.49\hsize}{!}{\includegraphics[angle=-90]{12724fg12d.ps}}
\caption{The $(u,v)$-plane coverage of four sources chosen to span the declination range 
of the TANAMI targets and thus representative of them. The short baselines near the 
center of each plot are produced by the telescopes in Australia. The long baselines are 
furnished by the Hartebeesthoek telescope in South Africa. The ``hole" in the $(u,v)$-plane 
coverage is a result of the absence of any telescopes between Australia and South Africa. 
Note that Hartebeesthoek is currently out of service and the long baselines are provided by 
O'Higgins in Antarctica and TIGO in Chile.}
\label{fig:uvplots}
\end{figure*}

Adequate $(u,v)$-coverage is a key consideration for any VLBI survey. It
is a particular concern when observing with the LBA since the LBA is an
ad hoc array and the locations of its constituent telescopes are not
ideal for producing uniform $(u,v)$-coverage. Representative plots of the
$(u,v)$-coverage for four sources with declinations spanning the range of
the TANAMI sample are shown in Fig.~\ref{fig:uvplots}. The shorter
baselines at the center of each plot are those between telescopes within
Australia. The long baselines at the periphery of each figure are those
to the Hartebeesthoek telescope in South Africa. The absence of
intermediate-length baselines between the shorter intra-Australian and
the trans-oceanic baselines remains the most important limiting factor on
image quality. However, as past imaging programs have found
\citep[e.g.,][]{Ojha2005}, this constraint does not preclude good images
provided special care is taken in both the calibration and imaging
process. Each epoch of observation included two sources which are
mutually visible to the LBA and VLBA. Our LBA images of these two sources
were checked for consistency with near-contemporaneous VLBA images and
revealed no problems.

\section{Notes on individual sources \label{sec:notesindiv}}
   
In this section we describe the morphology of each source after a brief
summary of its background and past observations where present. 
 
\paragraph{0047$-$579} This source is a bright high-redshift ($z=1.8$)
quasar from the radio-selected subsample, which has not yet been seen by
\textsl{Fermi} in its initial bright-source list \citep{AbdoLBAS2009}.
The 8.4\,GHz VLBI image by \citet{Ojha2005} shows a compact core with a
second component located $\sim$20\,mas to the west. Our TANAMI image
(Fig.~\ref{fig:tan_page1}) at the same frequency is at higher
resolution and shows a continuous well-collimated jet from the core
toward the outer component seen by \citet{Ojha2005}, which is partially
resolved by the TANAMI array. The tapered image
(Fig.~\ref{fig:tan_taper_page1}) reveals more extended emission along
the jet position angle out to $\sim$40\,mas from the core.

\paragraph{0208$-$512} This bright and highly polarized quasar
\citep{ImpeyTapia1988,ImpeyTapia1990} was a known EGRET source
\citep{Hartman1999} with its $\gamma$-ray spectrum being one of the
hardest detected AGN spectra in this energy range
\citep{vonMontigny1995,Chiang1995}. Right after the beginning of the
\textsl{Fermi} mission, 0208$-$512 was detected as a flaring $\gamma$-ray
source by the LAT \citep{Tosti2008} and it is also a member of the
3-month LAT bright source list sample \citep{AbdoLBAS2009}. 
\citet{Shen1998sl} find a lower limit for the Doppler factor of 10.2
using the ROSAT measurements (0.22\,$\mu$Jy at 1\,keV) of
\citet{Dondi1995}. From model fit parameters of the 1992 and 1993 data
they obtained a proper motion of 0.6$\pm$0.7\,mas/yr corresponding to a
transverse velocity of 16.8\,c (17$\pm$20\,c)\footnote{for $H_0=100$ and
$q_0=0.5$}. \citet{Tingay2002} find no significant detection of component
proper motion in this source. \citet{Shen1998} and \citet{Ojha2004} find
a slight extension of the compact core to the southwest. In the X-ray
regime \citet{Marshall2005} find a 4\,arcsec jet to the southwest in
addition to the core emission. Our image (Fig.~\ref{fig:tan_page1}) is
consistent but more sensitive showing a continuous well-collimated
twisting jet extending southwest to about $7$ milliarcseconds with
diffuse emission beyond that to over 20 milliarcseconds from the core.
Given the high fidelity of this image further TANAMI epochs should allow
us to pin down its expected superluminal motion.

\paragraph{0332$-$403} This BL\,Lac object has not been detected by EGRET
but was detected with \textsl{Fermi} between August and October 2008
\citep{AbdoLBAS2009}.  The source is very compact \citep{Ojha2004} with
previous reports of weak and short extensions to the east
\citep{Shen1998}, and to the west \citep{Fomalont2000}, respectively. Our
TANAMI image of this source has considerably lower resolution than most
of our other images because no trans-oceanic baselines were available and
does not resolve the source.


\paragraph{0405$-$385} This source is one of three unusually strong
intra-day variable (IDV) sources
\citep{Kedziora-Chudczer2006}\footnote{The other two strong IDV sources
are PKS 1257$-$326 and J1819+385}. It was not detected by EGRET but has
been detected, with low confidence, by the LAT \citep{AbdoLBAS2009}. This
is a very compact source but \citet{Zensus2002} find a component 1.5\,mas
to the west of the core. Similarly, \citet{Ojha2004} present a component
4\,mas west of their core. For the first-epoch TANAMI image of this
source, the absence of trans-oceanic baselines has resulted in lower
resolution, yielding a compact structure with an extension to the
west.

\paragraph{0438$-$436}
The VLBI image of this high-redshift quasar by \citet{Preston1989} shows two components, the core and an additional component 35\,mas to the southeast. A map with better resolution by \citet{Shen1998} resolves the component to the southeast of the core and reveals an additional component in between, separated from the core by about 7\,mas. \citet{Tingay2002} present a component which is located about 1\,mas to the eastsoutheast of the core. Our image has less resolution because only Australian antennas participated in this observation, but it does pick up more extended emission than previous images, revealing an extremely large continuous jet out to more than 50\,mas to the south-east.

\paragraph{0454$-$463} This flat spectrum radio source \citep{Kuehr1981}
is a highly polarized quasar. \citet{ImpeyTapia1990} measured a
polarization of 7.1\% and \citet{Wills1992} even 27.1\%. The source was
detected with EGRET \citep{Thompson1993} but somewhat surprisingly (given
that it is a very strong radio source with flux over 3.6\,Jy at 8.4\,GHz)
it has not been detected by the LAT as a bright $\gamma$-ray source in
the 3-month LAT data \citep{AbdoLBAS2009}. The VLBI image of 0454$-$463
by \citet{Ojha2004} shows a compact core without additional components.
Our image also shows a bright compact core with some very faint extended
emission out to $\sim$25\,mas to the east. The tapered image
(Fig.~\ref{fig:tan_taper_page1}) does show this eastern jet emission
more clearly but does not reveal any additional emission on larger
scales.

\paragraph{0506$-$612}
This source has been classified as a low-confidence potential
identification of the EGRET source 3EG\,J0512$-$6150 by \citet{Hartman1999}
and as a plausible identification by \citet{Mattox2001}. \citet{Ojha2004}
find an unresolved source. Our image (Fig.~\ref{fig:tan_page2}) reveals
a compact jet component at $\sim$2\,mas north-west of the core and a
fainter jet towards the west-north-west. Our tapered image in
Fig.~\ref{fig:tan_taper_page1} reveals more extended jet emission
turning to the west-south-west on scales $\sim$30--80\,mas from the core,
which was heavily resolved in the naturally-weighted image.

\paragraph{0518$-$458} Pictor A is one of the closest, powerful FR\,II
type radio galaxies \citep{FanaroffRiley1974}.  Its strong double-lobed
radio structure is oriented along the east-west direction 
\citep{Christiansen1977}. X-ray emission from the nucleus, the jet to the
west of the nucleus, the western radio hot spot, and the eastern radio
lobe was detected with Chandra \citep{Wilson2001}. \citet{Tingay2000}
found subluminal motions in the western jet. No parsec-scale counterjet
has been detected so far. Our image shows emission from the western jet
out to about 35\,mas from the core, with at least three compact jet
components in the inner 15\,mas, which may be associated with the
components C\,2, C\,3, and C\,4 seen by \citet{Tingay2000}, in which case
we would derive speeds of $\sim$0.15\,mas\,yr$^{-1}$ for all three
components. No counterjet emission is seen in the image shown in
Fig.~\ref{fig:tan_page2} , but we note that limitations in the
$(u,v)$-coverage do not allow us to put a strong upper limit on the
brightness of a potential counterjet.

\paragraph{0521$-$365} This source shows one of the best examples of an
optical synchrotron jet \citep{Danziger1979a,Boisson1989,Scarpa1999}. It
possesses strong extended radio and X-ray emission in addition to a
bright compact radio source and there are broad and variable nuclear
optical emission lines \citep{Ulrich1981,Scarpa1995}. The VLBI map of
this source at 5\,GHz by \citet{Shen1998} shows a core jet northwestward
consisting of a core and two additional components separated from the
core by $\sim$3.4\,mas and $\sim$8.3\,mas. High-resolution VLBI images
of this source have also been presented by \citet{Tingay1996b} and
\citet{Tingay2002_0521} but the TANAMI image shown in
Fig.~\ref{fig:tan_page2} shows the parsec-scale jet of this nearby
active galaxy with unprecedented resolution and image fidelity.

\paragraph{0537$-$441} This source has a GPS spectrum peaking at 5\,GHz
\citep{Tornikoski2001} and has been a strong variable EGRET gamma-ray
source \citep{Thompson1993_0537,Hartman1999}. It is known to be a strong 
source at all wavelengths \citep{Pian2002}.  It has been detected by the
LAT in a flaring state weeks after the beginning of \textsl{Fermi}
science operations \citep{Tosti2008}, and it is one of the two brightest
blazars in the southern $\gamma$-ray sky so far \citep{AbdoLBAS2009}.  A
VLA image by \citet{Cassaro1999} shows a curved jet-like structure
leading to the west. VLBI images at 2.3\,GHz and 8.4\,GHz for several
epochs are available in the United States Naval Observatory's Radio
Reference Frame Image Database at (RRFID; \url
{http://www.usno.navy.mil/RRFID/}).  At both frequencies the RRFID images
show a strong core with significant emission a few milliarcseconds to the
northeast. This is the same morphology as seen in our TANAMI image. We
find an unusually high brightness temperature exceeding $10^{14}$\,K of
the core which is both very compact (although slightly resolved) and very
bright. 

\paragraph{0625$-$354} This source exhibits a FR\,I radio-galaxy
morphology but its optical spectrum is more similar to a BL\,Lac object
\citep{Wills2004}.  The optical counterpart is a giant elliptical in the
center of the cluster Abell 3392 and exhibits a strong point source
nucleus \citep{Govoni2000}.  The VLBA map of this source by
\citet{Fomalont2000} shows a faint component to the southeast of the
core, which is consistent with the direction of the larger-scale jet. A
2.3\,GHz image obtained by \citet{Venturi2000} shows a diffuse outer jet
component at $\sim$200\,mas south-east of the core. Our tapered image
(Fig.~\ref{fig:tan_taper_page1}) shows jet emission on intermediate
scales out to $\sim$95\,mas from the core.  

\paragraph{0637$-$752} \citet{Yaqoob1998} find a peculiar emission line
in the X-ray spectrum of this radio-loud quasar with an energy of $1.60
\pm 0.07$\,keV and equivalent width $59^{+38}_{-34}$\,eV in the
quasar frame. 
This source was the first \textsl{Chandra} target, a 100\,kpc X-ray jet
to the west coinciding with the optical and radio contours was discovered
and at least 4 knots in the jet were resolved, which were separated from
the core by a few arcseconds \citep{Schwartz2000,Chartas2000}.  The VLBI
image by \citet{Tingay2002} shows a component about 1.5\,mas to the west
of the core. \citet{Ojha2004} find a component located 5\,mas from the
core.  \citet{Edwards2006} observed superluminal motion of this jet in
the parsec-scale. The TANAMI image confirms the structure seen in these
earlier images, revealing a jet out to about 10\,mas in the same
direction and significant components beyond that. The tapered image
(Fig.~\ref{fig:tan_taper_page1}) shows that the jet emission extends
beyond 50\,mas from the core. Despite being a strong superluminal blazar
this source was not detected by EGRET, and has not yet been detected by
the LAT \citep{AbdoLBAS2009}.

\paragraph{1104$-$445} The first southern VLBI Experiment observations
\citep{Preston1989} show a component located about 17\,mas to the
eastnortheast (position angle $75^\circ$) of the compact core of this
source. \citet{Shen1997} find a jet to the north-east with curvature from
the northeast to north at $\sim$1.8\,mas. \citet{Tingay2002} present a
highly resolved map, revealing structure in the northeast and southwest.
In the TANAMI image (Fig.~\ref{fig:tan_page3}), there is one jet
component $\sim$3\,mas north-east of the core, which is spread over
$>$8\,mas in the north-south direction. This structure may represent a
wide opening angle of the jet rather than curvature as suggested by
\citet{Shen1997}.

\paragraph{1144$-$379} This source has been classified as a BL Lac
object, due to optical, infrared and rapid radio variability, and its
featureless powerlaw spectrum \citep{Nicolson1979} but
\citet{VeronVeron2006} list it as a quasar. The VLBI map by
\citet{Shen1997} showed an unresolved core. Several RRFID epochs also
show an unresolved core at 8.4\,GHz but a few show an extension to the
southwest. The TANAMI image shows a clear jet in the same direction to
about 16\,mas along with significant emission at about 30\,mas from the
core better revealed in the tapered image
(Fig.~\ref{fig:tan_taper_page2}).  This bright, rapidly variable source
was not detected by EGRET but has already been detected by the LAT
\citep{AbdoLBAS2009}.

\paragraph{1257$-$326} This source is a flat-spectrum, radio-loud quasar,
which shows extreme intra-day variability due to interstellar
scintillation \citep{Bignall2003,Bignall2006}.  No VLBI image of this
source has been published so far. The TANAMI image shows a
well-collimated but possibly transversally resolved jet out to about
30\,mas north-west of the core. Because of its relatively low
signal-to-noise ratio, the core appears unresolved but the limit on its
brightness temperature is relatively low, only $10^{11}$\,K.

\paragraph{1313$-$333} The spectrum of this quasar is extremely flat and
the source, which is very variable at high radio frequencies, has been
associated with the EGRET source 3EG\,J1313-431
\citep{Nolan1996,Tornikoski2002}. VLBA images at 2.32 and 8.55\,GHz by
\citet{Fey1996} show jet components to the west, separated from the core
by 4.7\,mas at 2.32\,GHz, and 0.9 and 4.5\,mas at 8.55\,GHz,
respectively. Our image shows emission in the same direction and on the
same scales but at higher resolution than published before. 

\paragraph{1322$-$428} Centaurus A is the nearest AGN and is one of only
two radio galaxies detected by \textsl{Fermi} in its first three months
of observation \citep{AbdoLBAS2009}. This very well studied source was
also detected in $\gamma$-rays by EGRET. The TANAMI image shows a long
collimated jet extending to the northwest as well as a weak counterjet. 
The counterjet of Cen\,A was also seen in past VLBI images
\citep[e.g.,][]{Horiuchi2006}.

\paragraph{1323$-$526} This is a bright but relatively poorly studied
optically unclassified object. It shows intra-day variability
\citep{McCulloch2005}. \citet{Bignall2008} suggest a tentative
association of this flat-spectrum IDV radio source with the unidentified
EGRET source \mbox{3EG\,J1316$-$5244}. The TANAMI image shows a continuous
jet extending about 8\,mas to the south of the core.

\paragraph{1333$-$337}
(IC\,4296) A symmetric jet and counterjet system is visible in the kpc
regime, which is orientated in the northwest-southeast direction. The outer
lobes of the jets are separated by about 30\,arcmin
\citep{Goss1977,Killeen1986}. The central nuclear luminosity is
relatively weak compared to ``normal'' radio-loud AGN
\citep{Pellegrini2003}. This galaxy was not detected by EGRET, nor has it
been detected by the LAT in its first three months. The TANAMI image
shows its parsec-scale structure to have a jet and a counterjet aligned
northwest-southeast i.e. the same orientation as its kiloparsec
structure.

\paragraph{1424$-$418} Previous VLBI images of this highly optically
polarized quasar \citep{ImpeyTapia1988} showed jet components in
different directions. \citet{Preston1989} found a component to the
northeast separated by 23\,mas from the core, while the image by
\citet{Shen1998} showed a component about 3\,mas to the northwest of the
core. The VSOP image by \citet{Tingay2002} shows a northeastward
extension of the core and the one by \citet{Ojha2004} reveals weak
structure in the same direction. Our TANAMI image is in agreement with
the results of \citet{Preston1989,Tingay2002,Ojha2004} but reveals
substantially more detail than these previous images. The jet extends out
to about 40\,mas east from the core and is very diffuse and resolved,
best revealed in the tapered image in Fig.~\ref{fig:tan_taper_page2}.  

\paragraph{1454$-$354} This flat-spectrum radio quasar is not very well
studied. This source was included in our initial TANAMI sample as a
possible counterpart for the EGRET source \mbox{3EG\,J1500$-$3509}
\citep[with 1501$-$343 being an alternative
association;][]{Mattox2001,Sowards-Emmerd2004}.  This was the first
$\gamma$-ray blazar to be detected in outburst by \textsl{Fermi}
\citep{Marelli2008}. An analysis of the \textsl{Fermi} data from this
outburst are presented in \citet{LAT2009}, to which we contributed the
first deep high-resolution VLBI image of this source (naturally weighted
and tapered images reproduced in Fig.~\ref{fig:tan_page4} and
Fig.~\ref{fig:tan_taper_page2}).  These images show a bright compact core
with some diffuse emission extending northwest to about 40\,mas.

\paragraph{1501$-$343} Along with 1454$-$354, this source has been
suggested by \citet{Sowards-Emmerd2004} as a possible blazar counterpart
for 3EG\,J1500$-$3509. The early detection of 1454$-$354 by the LAT makes
1501$-$343 the more unlikely association of the 3EG source.
\citet{Petrov2007} did not detect this source at 22\,GHZ in six
observations with a minimum flux density limit of 170\,mJy. TANAMI
presents the first VLBI image of this source, revealing a very compact
structure with only a very weak possible extension to the south. Despite
the relatively low core flux density, the brightness temperature exceeds
$5\times10^{12}$\,K. 

\paragraph{1549$-$790} This luminous narrow-line radio galaxy has been
suggested to be an object at an early stage of its evolution
\citep{Tadhunter2001} and seems to be accreting at close to the Eddington
rate \citep{Holt2006} possibly related to a recent merger. Previously, a
one-sided core-jet type structure has been reported for this source
\citep{Holt2006}, which led to difficulties in understanding the absence
of broad emission lines and a strong non-stellar optical continuum, as
well as the presence of \textsc{hi} absorption \citep{Morganti2001}. The
tapered TANAMI image shown in Fig.~\ref{fig:tan_taper_page2} shows a
symmetric inner system with a jet and a pronounced counterjet extending
out to about 30\,mas east and west of the brightest feature, which we
tentatively identify as the core.  At larger distances towards the west,
there is very large and diffuse emission region ranging from
$\sim$50\,mas to $\sim$120--150\,mas from the core. There is an emission
gap of about 10\,mas between the inner and outer eastern structure and an
offset in declination may indicate jet curvature. Compared with the
images presented by \citet{Holt2006}, this image is much deeper. The
brightness distribution of the eastern outer structure is similar but not
the same as the one in \citet{Holt2006}. The difference could be due to
the better surface-brightness sensitivity in our image or due to
image-deconvolution uncertainties. The inner structure is shown in full
resolution (natural weighting) in Fig.~\ref{fig:tan_page4}. Immediately
to the west of the brightest central component, which we identify as the
core of the jet, there is a region of relatively low flux density. The
first bright feature in the counterjet seems to be located $\sim$4\,mas
to the west and may be interpreted as the base of the counterjet. The
central region between the two cores may then be attenuated in brightness
because of free-free absorption in a central obscuring torus similar to
the one in NGC\,1052 \citep[e.g.,][and references therein]{Kadler2004}.
This will be addressed in a future paper considering spectral information
from combined 8.4\,GHz and 22\,GHz TANAMI data.

\paragraph{1610$-$771} Using data with 22\,mas resolution
\citet{Preston1989} modeled this source as a 3.8\,Jy component 10\,mas in
extent with a 1.4\,Jy halo approximately 50\,mas in diameter. The VSOP
image by \citet{Tingay2002} revealed the small scale structure within
$\sim$1\,mas from the core to consist of 3 components at a position
angle of $-30$ degree with no obvious identification of the core. Our image
shows a curved jet towards the north-east, extending about 5\,mas from
the core and diffuse emission on larger scales to the north. The tapered
image in Fig.~\ref{fig:tan_taper_page2} shows this diffuse jet to extend
up to about 25\,mas north of the core.

\paragraph{1714$-$336} This is a possible counterpart of
3EG\,J1718$-$3313 \citep{Sowards-Emmerd2004} that has not yet been
detected by the LAT \citep{AbdoLBAS2009}. The TANAMI image of this BL Lac
appears to be the first at VLBI resolution and shows a bright core with a
jet extending over 20\,mas to the northeast. Only Australian antennas
participated in this observation so that the image resolution is worse
than for most other TANAMI images presented in this paper. The core
brightness temperature is unusually low.

\paragraph{1716$-$771} \citet{Tornikoski2002} suggested this unclassified
source of undetermined redshift as a possible counterpart for
3EG\,J1720$-$7820 but within the first three months of \textsl{Fermi}
all-sky observations, it did not show up as a bright $\gamma$-ray source
\citep{AbdoLBAS2009}. This is one of the faintest sources in the TANAMI
sample with no previous VLBI image. Fig.~\ref{fig:tan_page5} shows an
unresolved compact core of relatively high brightness temperature
$T_\textrm{B}>7\times10^{10}$. There is a faint jet extending about
4\,mas to the northeast.

\paragraph{1718$-$649} With a distance of 56\,Mpc this is one of the
closest and best studied GPS sources \citep[e.g.][]{Tingay2003b}.  Its
structure strengthens the assumption that GPS sources arise as a
consequence of galaxy merger activity \citep{Tingay1997}.  
\citet{TingayKool2003} suggest synchrotron self-absorption or free-free
absorption are the only possible processes responsible for the
gigahertz-peaked spectrum.  The maps by \citet{Tingay2002} and
\citet{Ojha2004} show two components separated by about 7\,mas. It was
not possible to identify which one corresponds to the core. The map by
\citet{Preston1989} also showed a double component aligned in the same
direction (southeast to northwest). Our TANAMI image shows two bright
components with similar alignment and separation but the northwestern
component is clearly more ``core-like" in appearance. There is also a
bright extension to the northeast of the ``core." We classify this source
as morphologically irregular. 

\paragraph{1733$-$565} This FR\,II radio galaxy has two extended lobes to
the southwest and northeast of the core, which are separated by
4.57\,arcmin \citep{Hunstead1982}, in between there is bridge and core
emission. \citet{Bryant2002} found rotating emission-line gas extended
perpendicular to the radio axis. The VLBI map by \citet{Ojha2004} shows a
compact core, without additional components. The TANAMI image shows a jet
as well as a counter jet aligned northeast and southwest i.e. in the same
direction as the large scale structure.

\paragraph{1759$-$396} This source is a possible counterpart of
3EG\,J1800$-$3955 \citep{Sowards-Emmerd2004} and is a low confidence
detection with the LAT \citep{AbdoLBAS2009}. The TANAMI image shows a
fairly typical core jet source with the jet extending northwest to about
10 mas.

\paragraph{1804$-$502} This source is a candidate counterpart for
3EG\,J1806$-$5005 \citep{Tornikoski2002} but it has not been a bright
$\gamma$-ray source during the first three months of \textsl{Fermi}
observations \citep{AbdoLBAS2009}. There is no high-resolution VLBI image
of this source in the literature. The TANAMI image shows an unresolved
high brightness-temperature core ($T_\textrm{B}>8.9\times10^{12}$\,K and
a weak jet to the north-west. 

\paragraph{1814$-$637} This is a compact steep spectrum (CSS) source with
three components oriented along a south-southeast to north-northwest
axis. \citet{Tzioumis2002} imaged this source at 2.3\,GHz and found two
strong components separated by more than 200\,mas and a weak
(15\,$\sigma$) component in between. \citet{Ojha2004} found an unresolved
core and an additional component about 90\,mas to the north-northwest at
8.4\,GHz but the alignment between these two images and the
identification of the core was not obvious. Our widefield tapered image
in Fig.~\ref{fig:tan_taper_page2} solves this problem. In total, we
detect three main components, of which the southernmost and the
northernmost can be identified by the outer structures seen by
\citet{Tzioumis2002}. Apparently, \citet{Ojha2004} did not pick up the
emission from the southern component. The central feature in our image
represents the core of the source with a parsec-scale jet extending to
the north, whose fine scale structure can best be seen in the naturally
weighted image in Fig.~\ref{fig:tan_page6}. The comparison with the
2.3\,GHz image by \citet{Tzioumis2002} suggests that the core component
may have an inverted spectrum and could be affected by free-free
absorption, although clearly a simultaneous multifrequency observation is
needed in order to address this quantitatively. Thus 1814$-$637 appears
to have a CSO morphology. Given the identification with a galaxy it is
likely that this is a relatively young radio source. This source was
considered as a possible EGRET detection by \citet{Edwards2005}.

\paragraph{1933$-$400} Based on VLA observations, \citet{Perley1982}
found that this source has a diffuse secondary component extending from
the core to 3.5\,arcsec at a position angle of $140^\circ$.
PKS\,1933$-$400 is identified with 3EG\,J1935$-$4022
\citep{Hartman1999,Edwards2005} but the source was not in the initial
\textsl{Fermi} 3-month bright source list. Our image shows a very linear
jet to the southeast, emanating from a bright core.


\paragraph{1954$-$388} This source has a high optical polarization, up to
11\% \citep{ImpeyTapia1988,ImpeyTapia1990} and a GPS-type spectrum
\citep{Tornikoski2001,Edwards2004}.
VLBI observations showed a compact core \citep{Preston1985,Shen1998}.
Based on VLBA observations, \citet{Fomalont2000} present a elongation of
the core to the southsouthwest, whereas \citet{Ojha2004} found a weak
component about 3\,mas to the west of the core. Our TANAMI image shows a
westward directed jet, extending about 5\,mas from the core. 

\paragraph{2005$-$489} This is one of the brightest known BL Lac sources
\citep{Wall1986} and it is classified as a high-frequency peaked BL Lac
(HBL) due to its X-ray-to-radio flux ratio \citep{Sambruna1995}. It is a
TeV source discovered by HESS, which has the softest VHE spectrum
($\Gamma=4.0$) ever measured from a BL Lac \citep{Aharonian2005_2005}.
The source was detected with EGRET \citep{Lin1999} and is one of the
bright $\gamma$-ray sources detected by \textsl{Fermi} in its first three
months of operations. The VLBI image by \citet{Shen1998} shows a compact
core. The image by \citet{Ojha2005} reveals an additional component about
3\,mas to the southwest of the core. Both our images in
Fig.~\ref{fig:tan_page6} and Fig.~\ref{fig:tan_taper_page3} show a low
surface brightness jet to the southwest of the core, which seems to have
a very wide opening angle.

\paragraph{2027$-$308} There is only very limited information on this
source in the literature. \citet{Grandi1983} note that it is probably a
member of the class of very-narrow-line emission galaxies.
\citet{Sowards-Emmerd2004} listed this source as a likely counterpart of
the EGRET source 3EG\,2034$-$3110.  Our TANAMI image shows an unresolved
core with a jet-like extension to the southwest.

\paragraph{2052$-$474} This source was identified with the EGRET source
3EG\,J2055$-$4716 \citep{Hartman1999}. Observations with Chandra did not
reveal extended X-ray emission, but there is a two-sided arcsecond-scale
radio jet in addition to the bright radio core \citep{Marshall2005}. The
VLBI image by \citet{Ojha2004} shows a compact core. Our image shows a
very weak jet to the west.

\paragraph{2106$-$413} The radio core of this quasar is moderately
polarized \citep[3.5\%,][]{ImpeyTapia1990}. ATCA observations indicate
that the radio spectrum peaks near 5\,GHz \citep{Kollgaard1995}.  The
VLBA image by \citet{Fomalont2000} shows a slightly elongated core. Our
image shows a bright component about 2--3\,mas east of the core, which
appears elongated in the north-south direction and additional diffuse jet
emission further to the east.

\paragraph{2149$-$306} This is a high-redshift, high-luminosity
radio-loud quasar with a strongly blueshifted Fe\,K$\alpha$ line at
$\sim$17\,keV in the quasar frame \citep{Yaqoob1999,Wang2003}. The VLBI
map at 8.4\,GHz by \citet{Ojha2005} shows a component to the west of the
core with a separation of 8.7\,mas. This component is not seen in our
TANAMI image, which rather shows jet emission within $\sim$5\,mas east of
the core. 

\paragraph{2152$-$699} This FR\,II radio source has a classic
double-lobed structure \citep{Fosbury1990}, and is one of the brightest
sources in the sky at 2.3\,GHz \citep{Wall1994}. \citet{Tadhunter1988}
find that the radio axis and optical emission line features on the
kiloparsec-scale are misaligned and suggest interaction between the radio
jet and an extra-nuclear cloud of gas. The parsec-scale radio jet aligns
strongly with optical emission line features \citep{Tingay1996a}. The
VSOP image by \citet{Tingay2002} shows a resolved core and highly linear,
narrow jet approximately 6\,mas to the northeast. \citet{Ojha2004} find a
similar morphology consisting of the core and jet component to the
northeast with separation of a few mas. Our image is in agreement with
these previous images of this source, revealing well-collimated but
knotty jet emission on intermediate scales out to about 30\,mas from the
core.

\paragraph{2155$-$304}
This is one of the brightest extragalactic X-ray sources in the sky and was detected with most high-energy satellites 
including EGRET \citep{Vestrand1995}. TeV $\gamma$-rays from this source
were observed \citep{Aharonian2005_2155} and a TeV flare occured in July
2006 \citep{Aharonian2007}. The VLBI map presented by \citet{Ojha2004}
shows a compact core without additional components. \citet{Piner2008}
could measure an apparent speed of a single jet component of
$0.93c\pm0.31c$ in a diffuse region $\sim$5\,mas southeast of the core.
Our TANAMI image does show the same structure on the same scales but at
somewhat higher dynamic range. In addition, there is a $\sim$15\,$\sigma$
component at the same position angle as the inner jet at $\sim 15$\,mas
from the core. The core itself has a relatively low brightness
temperature of $T_\textrm{B}=3.7\times10^{10}$\,K.

\paragraph{2204$-$540} This bright quasar is a member of our
radio-selected subsample and has a high polarization of $6.6\pm0.2$\%
\citep{Ricci2004}. There is only limited information on this source in
the literature, which is surprising for such a bright object. It has not
been seen by EGRET but it is among the brightest $\gamma$-ray sources
seen by \textsl{Fermi} in its first three months of observations
\citep{AbdoLBAS2009}. Our image shows a bright high
brightness-temperature core and a short and possibly curved jet in the
inner $\sim$4\,mas south and southwest of the core.

\paragraph{2326$-$477} \citet{Scott2004} find a compact, unresolved core
structure smaller than 0.1 mas$^2$ of 410\,mJy and a brightness
temperature above $3 \times 10^{11}$\,K at 5\,GHz. \citet{Tingay2003}
find a substantially higher mean flux-density at 5\,GHz of 1.63\,Jy with
ATCA and a moderate variability index of 0.05, suggesting that a
substantial fraction of the 5\,GHz total brightness is emitted on larger
scales unresolved for the space-VLBI array. Our image shows jet-like
emission peaking in two distinct components $\sim$6\,mas to the east and
$\sim$10\,mas to the north-east of the compact core at 8.4\,GHz.

\paragraph{2355$-$534} This is a optically violent and highly polarized
source \citep{ImpeyTapia1988,ImpeyTapia1990}. \citet{Shen1998} present a
VLBI image revealing a component to the southwest of the core with a
separation of 4.9\,mas. Our image shows at least two distinct jet
components along the same position angle, an inner one at $\sim$3.5\,mas,
which may consist of two subcomponents, and an outer one at $\sim
12$\,mas. No $\gamma$-ray detection of this object has been reported so
far.

\section{Discussion \label{sec:discussion}}

\subsection{Redshifts \label{subsec:redshifts}}
\begin{figure*}
   \centering
   \resizebox{\hsize}{!}{\includegraphics[angle=-90]{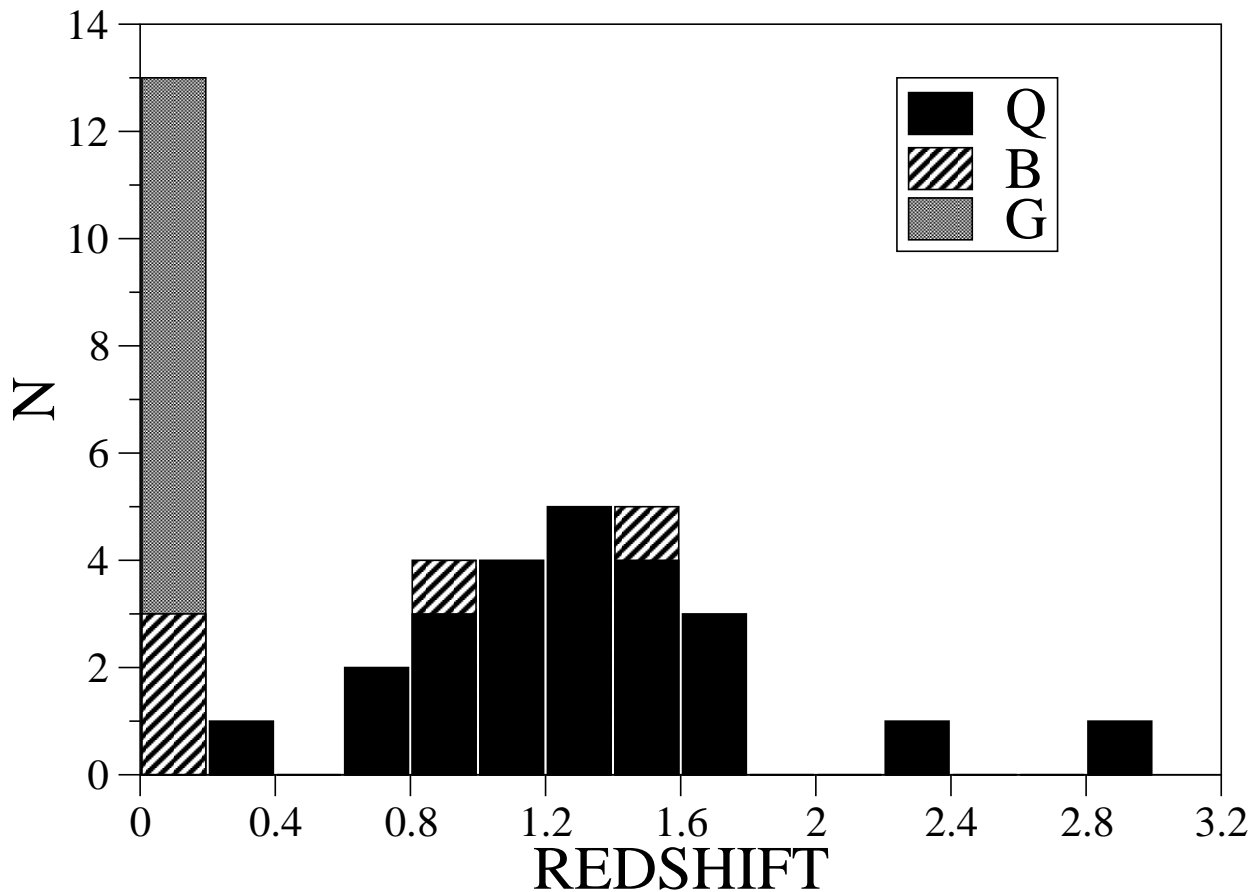}}
   \caption{Distribution of the redshifts of all TANAMI sources with optical identification shown.}
              \label{fig:Verteilungz}
    \end{figure*}

\begin{figure*}
   \centering
   \resizebox{\hsize}{!}{\includegraphics[angle=-90]{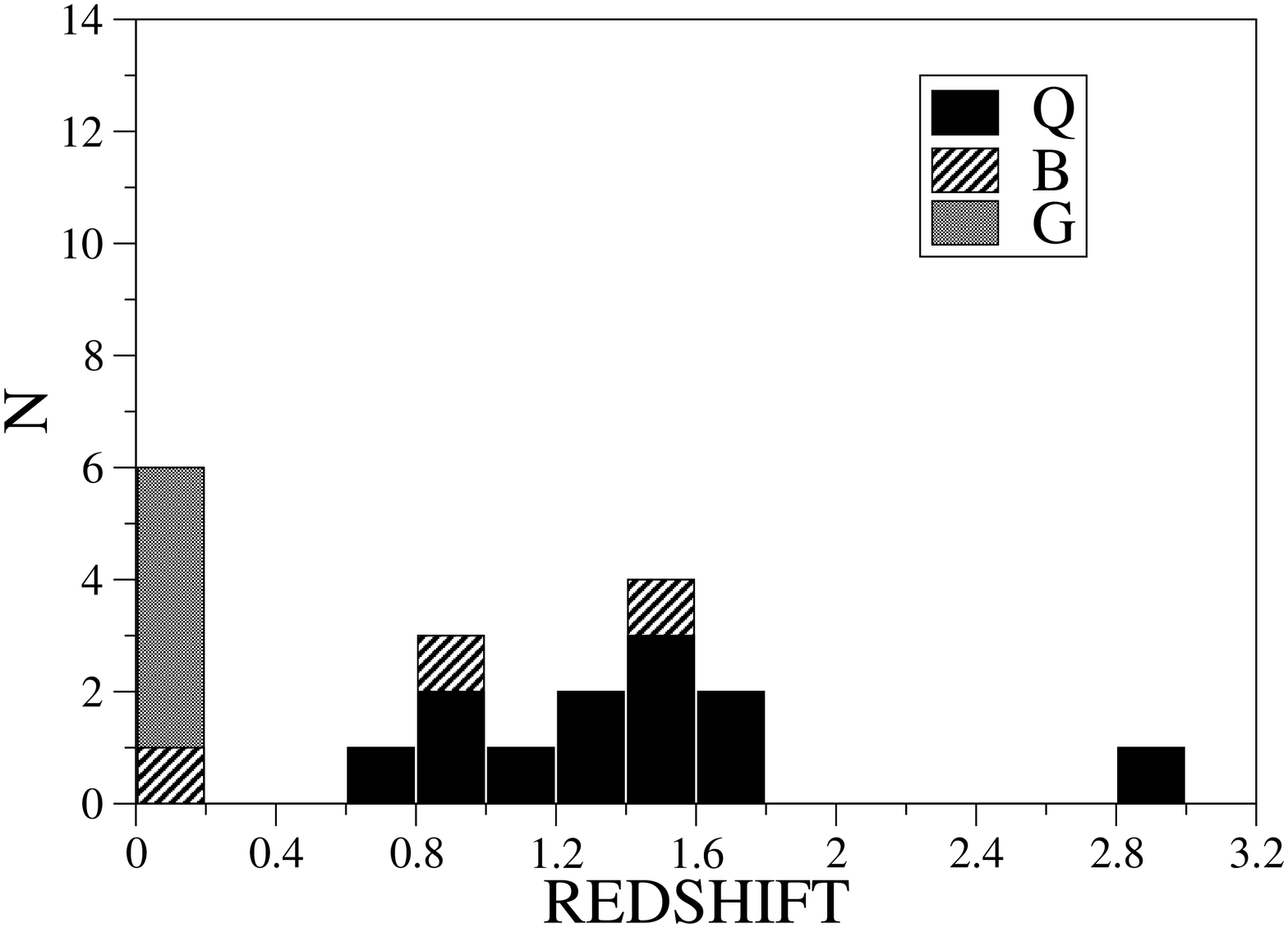}}
   \caption{Distribution of the redshifts of the radio-selected sub-sample of TANAMI sources.}
              \label{fig:VerteilungRadiosamplez}
    \end{figure*}

\begin{figure*}
   \centering
   \resizebox{\hsize}{!}{\includegraphics[angle=-90]{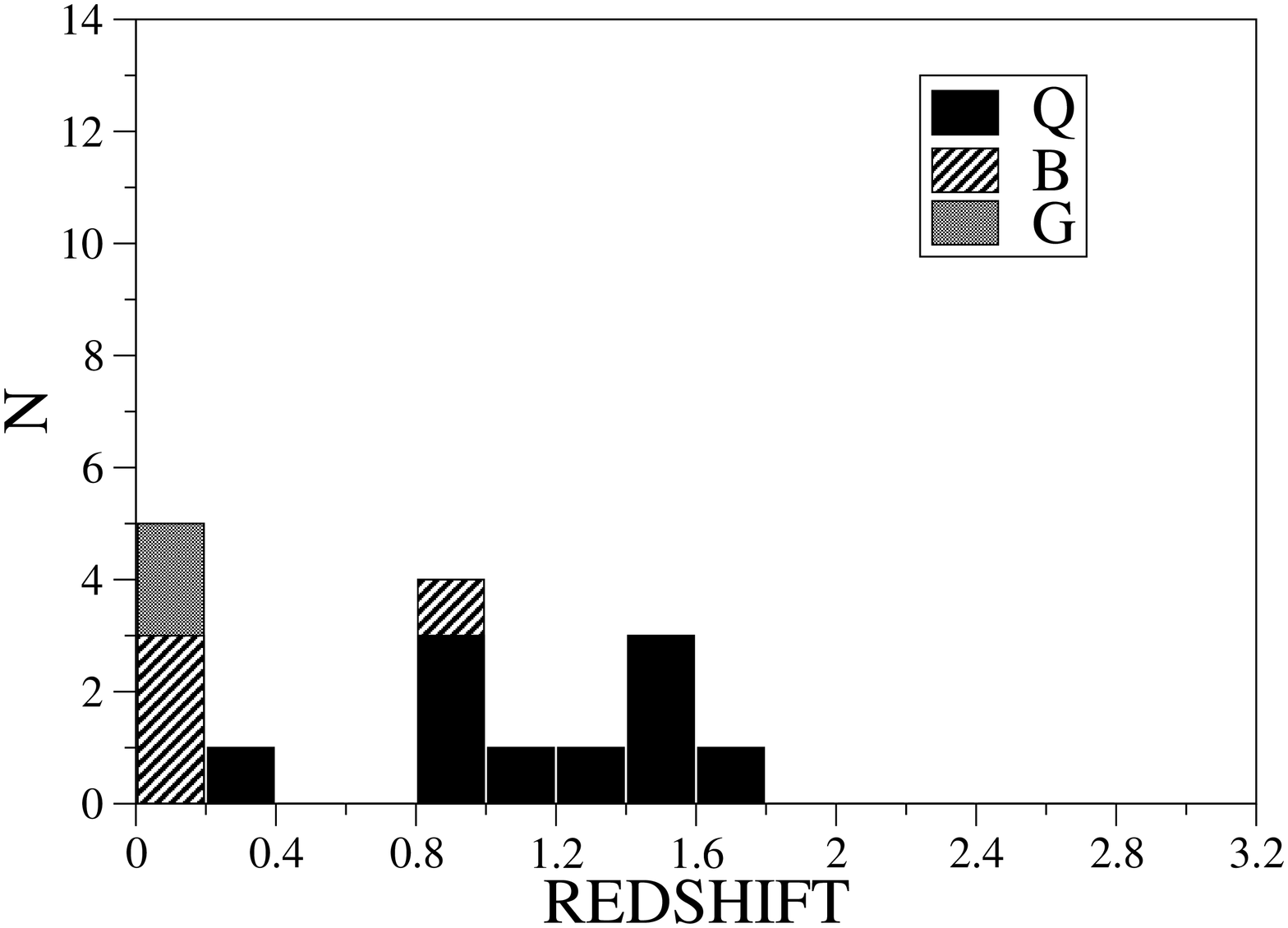}}
   \caption{Distribution of the redshifts of the EGRET-selected sub-sample of TANAMI sources. }
              \label{fig:VerteilungGammasamplez}
    \end{figure*}

\begin{figure*}
   \centering
   \resizebox{\hsize}{!}{\includegraphics[angle=-90]{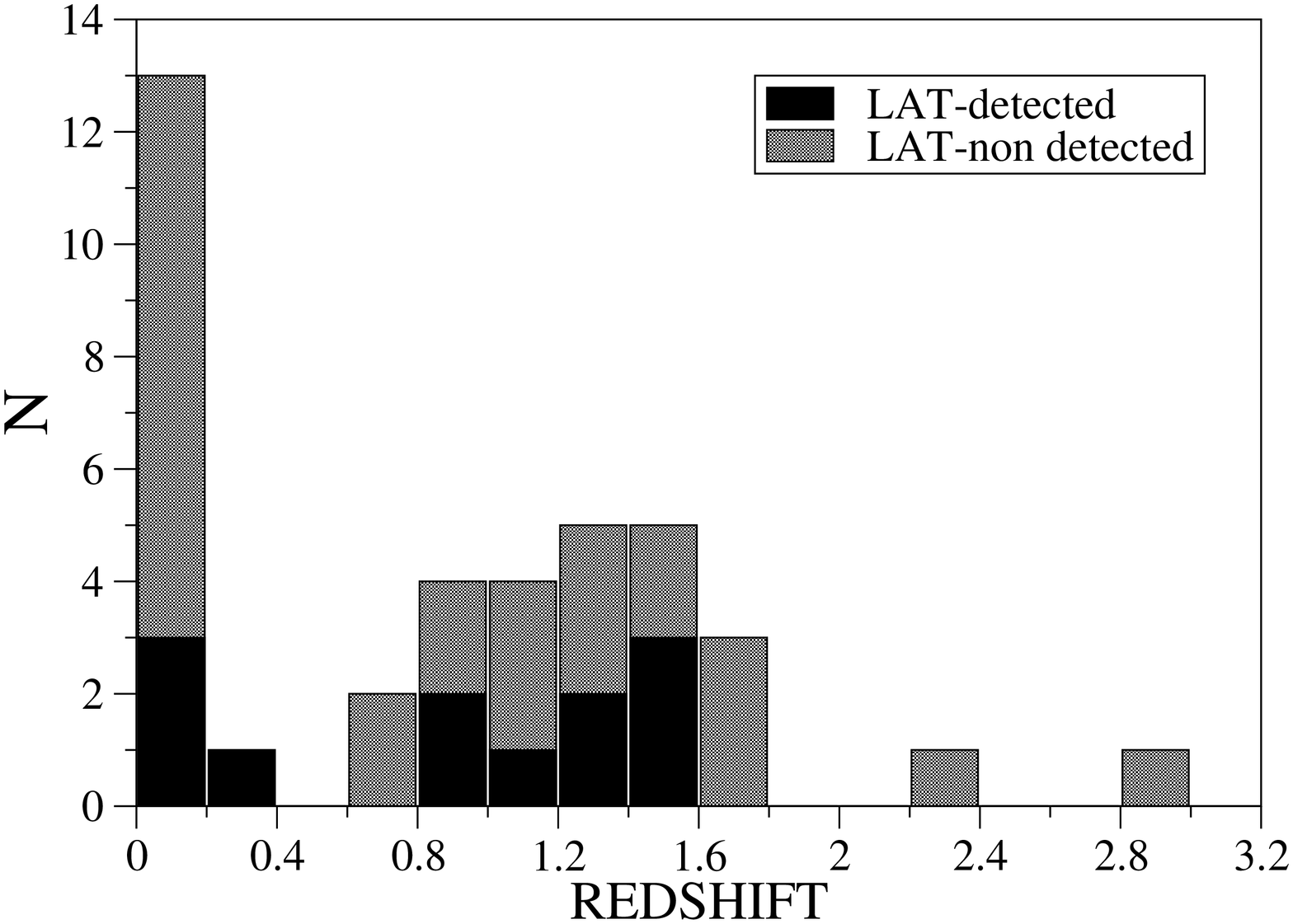}}
   \caption{Distribution of the redshifts of all TANAMI sources with LAT detections and non-detections shown. While broadly similar, most galaxies and none of the five most distant sources have been detected yet by the LAT.}
              \label{fig:VerteilungLATz}
    \end{figure*}

Figure~\ref{fig:Verteilungz} shows the redshift-distribution of all the
TANAMI sources.  For galaxies and BL Lac objects the distribution peaks
at $z<0.4$ while for quasars it peaks at $\sim 1.5$, with a maximum
redshift of $3$. This is similar to the distribution observed for the LAT
Bright AGN Sample \citep[LBAS;][]{AbdoLBAS2009} and for the EGRET blazars
\citep{Mukherjee1997}. The number of very low redshift AGN detected with
the \textsl{Fermi} LAT is expected to increase as fainter sources become
visible and more nearby radio galaxies are detected. The distributions of
the radio-selected and $\gamma$-ray selected sub-samples are shown
separately in Fig.~\ref{fig:VerteilungRadiosamplez} and
Fig.~\ref{fig:VerteilungGammasamplez}, respectively.  There is no
significant difference in the redshift distribution of these two
sub-samples which is consistent with the strong link between bright AGN
and $\gamma$-ray emission. Figure~\ref{fig:VerteilungLATz} again shows
the overall redshift distribution of all TANAMI sources but this time
with the LAT detections and non-detections indicated, based on the
3-month LAT bright-source list \citep{AbdoLBAS2009}. A comparison of
Fig.~\ref{fig:VerteilungGammasamplez} and
Fig.~\ref{fig:VerteilungLATz} shows that the redshift distribution of
EGRET detected AGN is similar to that of LAT detected AGN.
Figure~\ref{fig:VerteilungLATz} shows broadly similar redshift
distributions for the LAT detected and non-detected AGN in the TANAMI
sample with two interesting exceptions. First, most galaxies have not yet
been detected by the LAT. Rather more curiously, none of the five most
distant sources have been detected by the LAT.  As the size of the TANAMI
sample is increased, the interesting subsamples will become large enough
to allow exhaustive statistical tests which can better quantify the above
similarities and differences. 

\subsection{Luminosities \label{subsec:luminosity}}
For all 38 TANAMI sources that have a published redshift, the core and
the total luminosity was calculated assuming isotropic emission.
The results are shown in the final two columns of
Table~\ref{table:sourcestructure}. For both LAT detected and non-detected
sources, the values range from about $10^{22}$ to almost $10^{29}$
$\mathrm{W\,Hz^{-1}}$ with two thirds of both of these categories of
sources having total luminosity above $10^{27}$ $\mathrm{W\,Hz^{-1}}$.
There is a clear difference between the distribution of luminosity of
different optical types with all eight galaxy luminosities below
$10^{26}$ $\mathrm{W\,Hz^{-1}}$ and all twenty-four quasar luminosities
above $10^{26}$ $\mathrm{W\,Hz^{-1}}$ (all but one above $10^{27}$
$\mathrm{W\,Hz^{-1}}$). The five BLLacs in the sample are evenly
distributed between $10^{24}$ and $10^{28}$ $\mathrm{W\,Hz^{-1}}$. 

None of the five most luminous sources (which are also the five most
distant sources, see \ref{subsec:redshifts}) have been detected by the
LAT. More intriguing, none of the nine most luminous jets (obtained by
taking the difference of total and core luminosity) are detected which,
taken at face value, would suggest an unexpected anti-correlation between
jet luminosity and $\gamma$-ray brightness. More typically, the four most
luminous BLLac sources have been detected by the LAT. 

\subsection{Morphology and Connection to \textsl{Fermi} \label{subsec:morphology}}
To discuss the morphology of TANAMI sources we have adopted the
classification scheme used by \citet{Kellermann1998} which places objects
into four categories. Sources that appear barely resolved are considered
``compact" (C), those with the most compact component at either end of
the image are considered ``single-sided" (SS) and those with the most
compact component in the middle of the image are considered
``double-sided" (DS). Finally, there is a category of sources with
``irregular" (Irr) structure which includes sources with morphology that 
does not fall into the first three categories. This scheme has the virtue
of not making any assumptions about the physical nature of the objects
under study, separating the $\it description$ of the observations from
their $\it interpretation$. For an excellent discussion of the possible 
physical conditions associated with these categories see Sect.~4 of 
\citet{Kellermann1998}. 

Three TANAMI sources, 0332$-$403, 0405$-$385, and 1501$-$343, appear
compact from our images. However, we only classify 1501$-$343 as
``compact" since our images for the first two sources have a lower
resolution due to the absence of trans-oceanic baselines and past images
indicate the presence of some weak extended structure. Optically
1501$-$343 is an ``unclassified" source.  1322$-$428, 1333$-$337,
1549$-$790, 1733$-$566 and 1814$-$637 are double-sided. All five are
galaxies. Another galaxy, 1718$-$649 is the only ``irregular" source in
our sample.  All of the remaining 36 sources ($84\%$ of the TANAMI
sample) are single-sided (Table~\ref{table:sourcedistribution}). These
include all quasars, all BL Lacs, four of the ten galaxies and two of the
three optically unidentified sources in the TANAMI sample.
Characteristics of individual sources are discussed in
Sect.~\ref{sec:notesindiv}. 

Since the $\gamma$-ray emission is likely to be beamed and orientation
dependent, we would expect to find differences in the parsec scale
morphology of $\gamma$-loud and $\gamma$-quiet objects. However, past
comparisons of VLBI morphologies of EGRET detected sources with those not
detected by EGRET failed to show any connection between observed
$\gamma$-ray emission and parsec scale structure \citep{Taylor2007}. This
was attributed to the fact that nearly all EGRET sources were detected
only when flaring and that most sources lay within a factor of 10 of
EGRET's minimum detectable flux.  The expectation is that essentially all
bright compact radio sources will be $\gamma$-ray loud if observed with
significantly better sensitivity than EGRET. 

After three months of observations \textsl{Fermi} found 12 of our 43
TANAMI sources ($28\%$) to be bright $\gamma$-ray sources with $> 10
\sigma$ significance. This includes, 11 of the 36 SS sources. However,
only one (Centaurus A) of the 5 DS sources have been detected so far. In
terms of optical identification, 11 of the \textsl{Fermi} detections are
quasars and BL Lacs with Cen A the sole galaxy detected. This is
consistent with the canonical picture that the ``core-jet" nature of
quasars and BL Lacs is the result of differential Doppler boosting of an
intrinsically symmetric twin-jet. That such AGN are associated with
$\gamma$-ray emission was a key finding of EGRET and is consistent with
theoretical models. None of the three optically unidentified sources in
our sample have been detected so far. 

Interestingly, two of the four known intraday variable (IDV) sources in
our sample (0405$-$385 and 1144$-$379) have been detected by \textsl{Fermi}.
All four IDV sources have a `SS' morphology. This is consistent with our
current understanding of IDV sources as being among the most extreme
members of the blazar class. 

To compare the opening angles of \textsl{Fermi}-detected and non-detected
blazar jets, we fitted circular Gaussian components to the visibility
data and measured the angle at which the innermost jet component appears
from the position of the core of the jet
(Table~\ref{table:openingangle}). In two cases (0332$-$403 and 0405$-$385)
the lack of long baselines did not allow us to resolve the inner jet
structure and to model-fit any jet components. In one case (0454$-$463),
the jet was too weak and partially resolved, and in one case (1501$-$343),
the source was unresolved even on the longest baselines. We excluded all
galaxies from this analysis, first, because we cannot unambiguously
determine the core position in a number of galaxy jet-counterjet systems
and, second, because galaxies are usually not detected by \textsl{Fermi}.
Our initial analysis suggests that 7 out of 9 ($78\%$) LBAS sources have
an opening angle $>30$ degrees while only 4 out of 15 ($27\%$) non-LBAS
sources have an opening angle $> 30$ degrees. 
This would suggest that $\gamma$-ray bright jets have either smaller
Lorentz factors (the width of the relativistic beaming cone is
$\sim$$1/\Gamma$) or that they are pointed closer to the line of sight
than $\gamma$-ray faint jets. An inverse correlation between $\gamma$-ray
brightness and the beaming seems very unlikely and, in fact, previous
studies have shown that the Lorentz factor of $\gamma$-ray brighter jets
as seen with \textsl{Fermi} are higher than for $\gamma$-ray fainter jets
\citep[e.g.,][]{Lister2009a,Kovalev2009}. If confirmed by future analysis
of larger samples, the result that $\gamma$-ray brighter jets have larger
observed opening angles might thus imply that these appear geometrically
increased in projection because of smaller angles to the line of sight.

\subsection{Brightness Temperature \label{subsec:t_b}}
We identified the core component for each source based on its morphology
and, in some cases, past VLBI images and identifications. In all cases,
this approach led to the identification of the core as the brightest and
most compact feature, either at the end of a one-sided jet or at the
center of a double-sided twin-jet. Gaussian models were fitted to the
core visibility data using {\sc difmap}. We did this by replacing the
{\sc clean} components from the core region in the final model with an
elliptical Gaussian component. The brightness temperature of the fitted
core component was then calculated in the rest frame of the source using
the expression:
\begin{equation}
T_\textrm{B}=\frac{2 \ln{2}}{\pi k} \frac{S_\textrm{core} \lambda^2 (1+z)}{\theta_\textrm{maj} \theta_\textrm{min}}
\end{equation}
where $S_\textrm{core}$ is the flux density of the core and
$\theta_\textrm{maj}$ and $\theta_\textrm{min}$ are the major and minor
axis FWHMs, respectively, of the Gaussian component. $k$ is the Boltzmann
constant, $z$ the redshift and $\lambda$ the wavelength of observation
\citep[compare][]{Kovalev2005}.

In order to distinguish between cores that were resolved and those that
were not we compared the fitted component sizes $\theta_\textrm{fit}$ with the
theoretical resolution limit of our array $\theta_\textrm{lim}$. We
calculated $\theta_\textrm{lim}$ taking into account the synthesized
beamsize, $\theta_\textrm{beam}$ (where $\theta_\textrm{beam}$ is the
geometric mean of the major and minor beam axes) and the signal-to-noise
ratio (SNR) of the core component following \citet[][Eq.~2]{Kovalev2005}.
Components with fitted sizes $\theta_\textrm{fit} < \theta_\textrm{lim}$
are considered unresolved and we determine lower limits of their
calculated brightness temperatures by using $\theta_\textrm{lim}$ as an
upper limit on their size. It should be noted that all other brightness
temperatures are limits in a similar sense, because VLBI observations at
a given frequency can never rule out the possibility that even smaller
structures inside the cores are dominating the core emission.

   \begin{figure*}
   \centering
   \resizebox{\hsize}{!}{\includegraphics[angle=0]{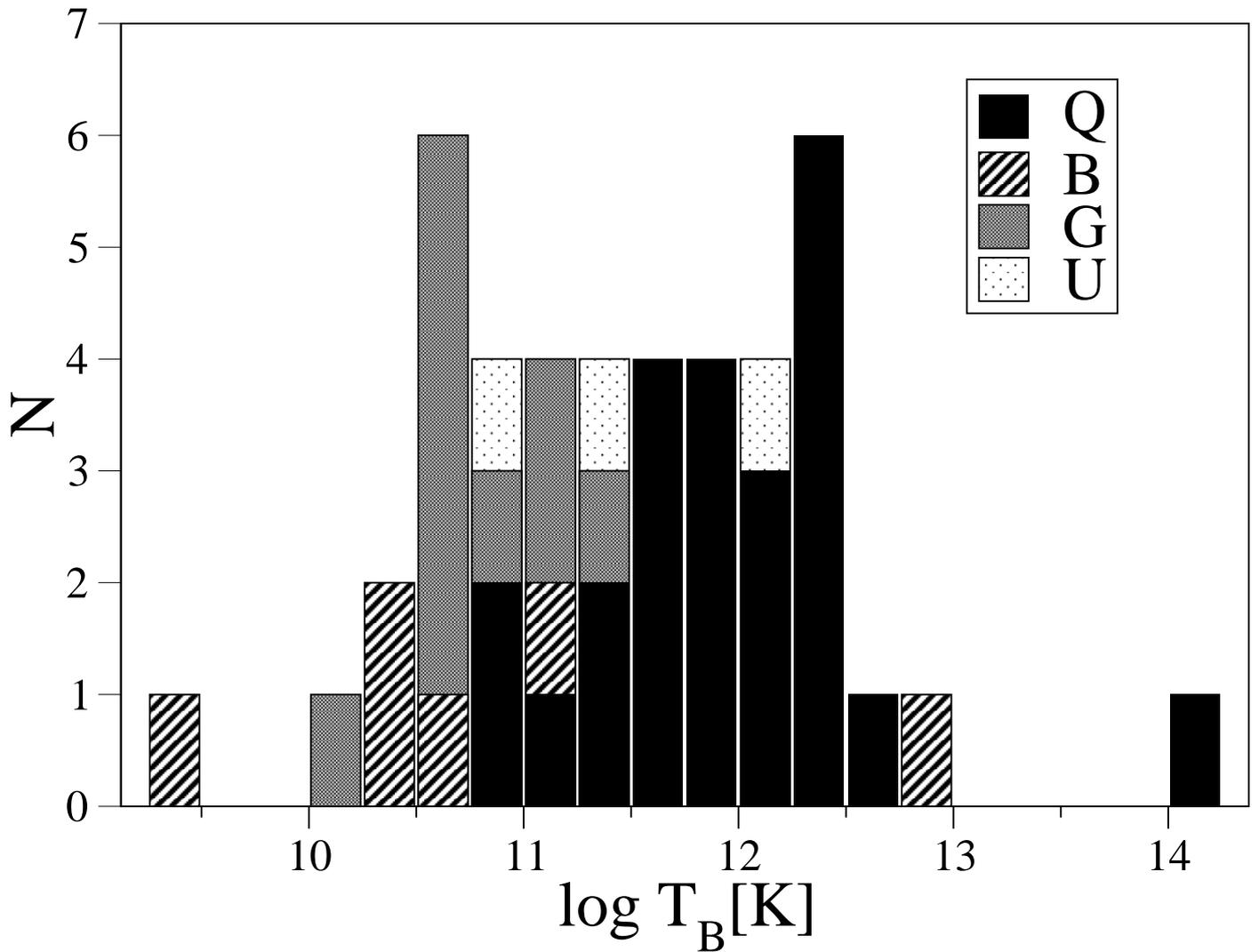}}
   \caption{Distribution of core brightness temperature of all TANAMI sources with optical identification shown.}
              \label{fig:VerteilungTB}
    \end{figure*}
    
   \begin{figure*}
   \centering
   \resizebox{\hsize}{!}{\includegraphics[angle=0]{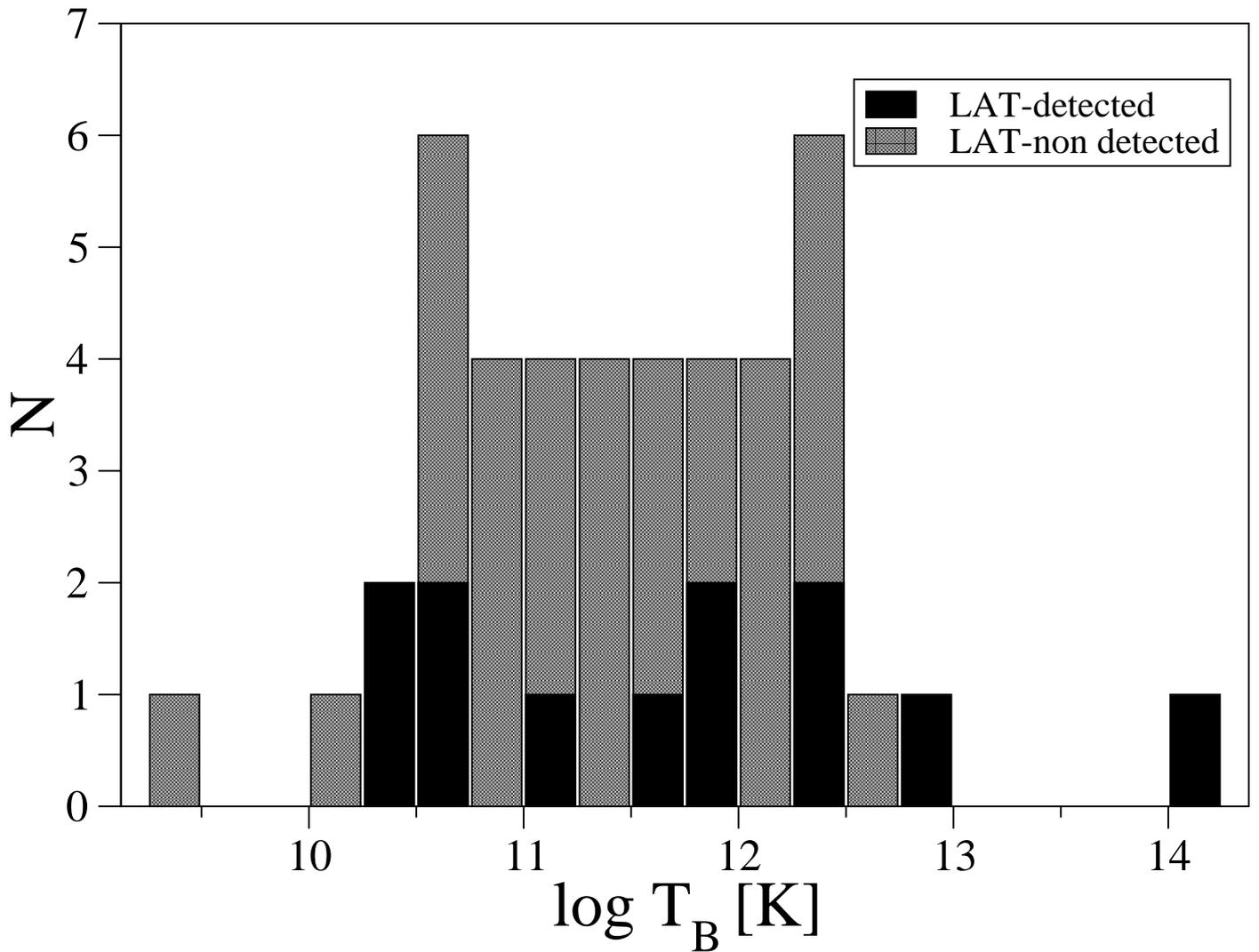}}
   \caption{Distribution of core brightness temperature of all TANAMI sources with LAT detections and non-detections shown.}
              \label{fig:VerteilungLATTB}
    \end{figure*}

The core brightness temperatures for all sources are listed in
Table~\ref{table:sourcestructure} and their distribution is shown in
Fig~\ref{fig:VerteilungTB}. The higher end of the distribution is
dominated by quasars while the lowest brightness temperatures are those
of BL Lacs and galaxies.  We find 2 sources with fully unresolved cores
(1501$-$343, and 2027$-$308) and 2 which are unresolved perpendicular to
the jet axis and resolved along it (1313$-$333 and 1804$-$502). Fourteen 
sources show a maximum brightness temperature below the
equipartition value of $10^{11}$ K \citep{Readhead1994} with thirty
sources having values below the inverse Compton limit of $10^{12}$\,K
described by \citet{Kellermann1969}. As many as thirteen sources
exceed this limit, one of them substantially (0537$-$441). This is most
likely a result of Doppler boosting that is commonly seen in blazars.
However, contributions from exotic mechanisms such as coherent emission
and relativistic proton emission and/or non-simple geometries cannot be
ruled out \citep{Kellermann2004}. The median value is near {\bf
$2.1\times10^{11}$}K with the maximum value exceeding $10^{14}$ K. This
is comparable to the results of the MOJAVE sample \citep{Kovalev2005}.

Fig~\ref{fig:VerteilungLATTB} shows the core brightness temperatures of
LAT detections and non-detections, based on the 3-month LAT list.  There
does not appear to be any significant difference between LAT detections
and non-detections. It is interesting to note that many sources with high
brightness temperatures, and thus expected to have high Doppler factors,
remain undetected by the LAT. Nine of the thirteen with values above 
$10^{12}$ K remain undetected.

\section{Conclusions \label{sec:conclusions}}

Our first epoch 8.4\,GHz TANAMI images show that the Australian Long
Baseline Array and associated telescopes provide high quality images of
$\gamma$-ray blazars that can be used to study the physics of blazars. In
many cases these images represent a substantial improvement on published
work. The addition of telescopes in Antarctica and Chile is expected to
compensate for the loss of the Hartebeesthoek telescope in South Africa
and to improve the current $(u,v)$-coverage. 

The TANAMI sample has been defined as a hybrid radio and $\gamma$-ray
selected sample of AGN south of $\delta = -30^{\circ}$. Of this sample,
$84\%$ of sources show a one-sided morphology, $12\%$ (all galaxies) are
double sided, while one optically unidentified source is compact and one
galaxy has a irregular morphology. Of these quasars and BL Lacs have
similar morphologies, all being single-sided. The ten galaxies in the
sample include five double sided objects, one irregular object and four
single sided objects.  

About 28\,\% of TANAMI sources have been detected by the \textsl{Fermi}
LAT after three months of observations. When galaxies are excluded,
initial analysis shows that $78\%$ of sources detected by the LAT have
opening angles $> 30$ degrees compared to just $27\%$ of non-LBAS
sources. This suggests that $\gamma$-ray bright jets are pointed closer
to the line of sight than $\gamma$-ray faint jets, with the observed
opening angles appearing geometrically increased. This result should be
regarded as preliminary owing to the modest number of sources available
for analysis at this time.

The redshift distribution of the BL Lacs and quasars in the TANAMI sample
is similar to the distributions seen for the LBAS and EGRET blazars. No
difference is seen between the radio- and $\gamma$-ray selected
subsamples. The redshift distributions of EGRET and LAT detected AGN are
also similar. However, most galaxies and none of the five most distant
and most luminous sources have been detected by the LAT. Galaxies have
the lowest luminosities, BLLacs are more luminous (than galaxies) as a
group while quasars dominate the high end of the luminosity distribution.
None of the nine most luminous jets have been detected by the LAT so far.

The high end of the brightness temperature distribution is dominated by
quasars with the lower end composed mostly of BL Lacs and galaxies.
Thirteen sources have a maximum brightness temperature below the
equipartition value, and 29 sources have a value below the inverse
Compton limit putting about a third of the values above this limit.  The
median value is near $3\times10^{11}$K with the maximum value exceeding
$10^{14}$ K. There does not appear to be any significant difference in
the brightness temperature distributions of LAT detections and
non-detections. Many of the sources with very high brightness
temperatures have not yet been detected by the LAT. 
      
The initial results presented here will be augmented by 22\,GHz images as
well as multi-epoch data in future papers to address our scientific
questions including those that require spectral and kinematic
information. Of particular, interest will be TANAMI epochs observed since
the launch of \textsl{Fermi} satellite thus providing near-simultaneous
data at radio, $\gamma$-ray as well as at intermediate wavelengths from
coordinated observations with other instruments. As discussed in
Sect.~\ref{sec:intro}, TANAMI and \textsl{Fermi} data will also be used
with neutrino data from ANTARES and KM3NET for the identification and
study of neutrino point sources.


\begin{acknowledgements}
We are grateful to Dirk Behrend, Neil Gehrels, Julie McEnery, David
Murphy, and John Reynolds, who contributed in numerous ways to the
success of the TANAMI program so far. Furthermore, we thank the
\textsl{Fermi}/LAT AGN group for the good collaboration. 

The Long Baseline Array is part of the Australia Telescope which is
funded by the Commonwealth of Australia for operation as a National
Facility managed by CSIRO. This work made use of the Swinburne University
of Technology software correlator, developed as part of the Australian
Major National Research Facilities Programme and operated under licence.
M. K. has been supported in part by an appointment to the NASA
Postdoctoral Program at the Goddard Space Flight Center, administered by
Oak Ridge Associated Universities through a contract with NASA. We would
like to thank the staff of the Swinburne correlator for their unflagging
support.  This research has made use of data from the NASA/IPAC
Extragalactic Database (NED, operated by the Jet Propulsion Laboratory,
California Institute of Technology, under contract with the National
Aeronautics and Space Administration); and the SIMBAD database (operated
at CDS, Strasbourg, France). This research has made use of NASA's
Astrophysics Data System. This research has made use of the United States
Naval Observatory (USNO) Radio Reference Frame Image Database (RRFID).

\end{acknowledgements}

\bibliographystyle{jwaabib}
\bibliography{mnemonic,aa_abbrv,tanami}

\clearpage

\begin{deluxetable}{lrl}
\tabletypesize{\scriptsize}
\tablecaption{The Long Baseline Array and Affiliated Telescopes\label{table:antennas}}
\tablewidth{0pt}
\tablehead{
\colhead{Telescope} & \colhead{Diameter} & \colhead{Location} \\
\colhead{} & \colhead{(meters)}  & \colhead{}
}
\startdata
Parkes & 64 &  Parkes, New South Wales, Australia \\
ATCA & 5x22 & Narrabri, New South Wales, Australia \\
Mopra & 22  & Coonabarabran, New South Wales, Australia \\
Hobart & 26  & Mt. Pleasant, Tasmania, Australia \\
Ceduna & 30 & Ceduna, South Australia, Australia \\
Hartebeesthoek\tablenotemark{a} & 26 & Hartebeesthoek, South Africa \\
DSS43\tablenotemark{b} & 70 & Tidbinbilla, ACT, Australia \\
DSS45\tablenotemark{b} & 34 & Tidbinbilla, ACT, Australia \\
O'Higgins\tablenotemark{c} & 9 & O'Higgins, Antarctica \\
TIGO\tablenotemark{c} & 6 & Concepcion, Chile \\
\enddata
\tablenotetext{a}{Not available since September 2008}
\tablenotetext{b}{Operated by the Deep Space Network of the National Aeronautics and Space 
Administration}
\tablenotetext{c}{Operated by Bundesamt fŸr Kartographie und GeodŠsie (BKG) \url{http://www.bkg.bund.de/nn_147094/EN/Home/homepage__node.html__nnn=true}}

\end{deluxetable}

\clearpage

\begin{deluxetable}{lll}
\tabletypesize{\scriptsize}
\tablecaption{Summary of Observations\label{table:epochs}}
\tablewidth{0pt}
\tablehead{
\colhead{Epoch} & \colhead{Participating Telescopes} & \colhead{Remarks} \\
}
\startdata
2007NOV10 &  Parkes, ATCA, Mopra, Hobart, Ceduna, Hartebeesthoek &  \ldots \\
2008FEB07 & Parkes, ATCA, Mopra, Hobart, Ceduna, Hartebeesthoek, DSS43 &  Hart for 15 and DSS43 for 5 hours only \\
2008MAR28 &  Parkes, ATCA, Mopra, Hobart, Ceduna, Hartebeesthoek, DSS45 &  DSS45 intermittent \\
2008JUN09 &  Parkes, ATCA, Mopra, Hobart, Ceduna, Hartebeesthoek &  \ldots \\
\enddata
\end{deluxetable}

\clearpage

\begin{deluxetable}{ccccccccccc}
\tabletypesize{\tiny}
\tablewidth{0pt}
\tablecaption{Source List \label{table:sourcelist}}
\tablehead{
\colhead{Source\tablenotemark{a}} & \colhead{Name\tablenotemark{b}} & \colhead{R.A.} & \colhead{Decl.} & \colhead{ID\tablenotemark{d}} & \colhead{Magnitude} & \colhead{Redshift} & \colhead{Radio}& \colhead{$\gamma$}  &\colhead{EGRET\tablenotemark{g}} & \colhead{LAT\tablenotemark{h}} \\
\colhead{} & \colhead{} & \colhead{(J2000.0)\tablenotemark{c}} & \colhead{(J2000.0)\tablenotemark{c}} & \colhead{} & \colhead{$V$\tablenotemark{e}} & \colhead{$z$\tablenotemark{f}} & \colhead{Sample} & \colhead{Sample} &\colhead{} & \colhead{}
}
\startdata
0047$-$579&                         & $00^\mathrm {h}49^\mathrm {m}59\rlap{.}^\mathrm {s}4731$ & $-57^\circ 38^\prime 27\rlap{.}^{\prime\prime}339$ & Q & 18.50 & 1.797$^{1}$   & Y & N & N & N\\
0208$-$512&                         & $02^\mathrm {h}10^\mathrm {m}46\rlap{.}^\mathrm {s}2004$ & $-51^\circ01^\prime02\rlap{.}^{\prime\prime}891$ & B & 16.93 & 0.99$^{2}$      & Y & Y & Y & Y \\
0332$-$403&                         & $03^\mathrm {h}34^\mathrm {m}13\rlap{.}^\mathrm {s}6544$ & $-40^\circ08^\prime25\rlap{.}^{\prime\prime}396$ & B & 18.50 & 1.445$^{3}$     & Y & N & N & Y \\
0405$-$385&                        & $04^\mathrm {h}06^\mathrm {m}59\rlap{.}^\mathrm {s}0353$ & $-38^\circ26^\prime28\rlap{.}^{\prime\prime}042$ & Q & 18.00 & 1.285$^{4}$      & N & N & N & Y* \\
0438$-$436&                        & $04^\mathrm {h}40^\mathrm {m}17\rlap{.}^\mathrm {s}1799$ & $-43^\circ33^\prime08\rlap{.}^{\prime\prime}602$ & Q & 18.8 & 2.863$^{5}$       & Y & N & N & N \\
0454$-$463&                         & $04^\mathrm {h}55^\mathrm {m}50\rlap{.}^\mathrm {s}7724$ & $-46^\circ15^\prime58\rlap{.}^{\prime\prime}681$ & Q & 17.40 & 0.8528$^{6}$    & Y & Y & Y & N \\
0506$-$612&                         & $05^\mathrm {h}06^\mathrm {m}43\rlap{.}^\mathrm {s}9887$ & $-61^\circ09^\prime40\rlap{.}^{\prime\prime}993$ & Q & 16.85 & 1.093$^{3}$     & N & Y & C & N \\
0518$-$458&    Pictor A       & $05^\mathrm {h}19^\mathrm {m}49\rlap{.}^\mathrm {s}69$ & $-45^\circ46^\prime44\rlap{.}^{\prime\prime}5$ & G & 16.45 & 0.035058$^{7}$            & N & N & N & N \\ 
0521$-$365&   ESO 362-G 021 & $05^\mathrm {h}22^\mathrm {m}57\rlap{.}^\mathrm {s}9846$ & $-36^\circ27^\prime30\rlap{.}^{\prime\prime}848$ & B & 14.50 & 0.055338$^{8}$        & Y & Y & C & N \\
0537$-$441&                        & $05^\mathrm {h}38^\mathrm {m}50\rlap{.}^\mathrm {s}3614$ & $-44^\circ05^\prime08\rlap{.}^{\prime\prime}934$ & Q & 15.50 & 0.894$^{1}$    & Y & Y & Y & Y \\
0625$-$354&     OH-342     & $06^\mathrm {h}27^\mathrm {m}06\rlap{.}^\mathrm {s}72$ & $-35^\circ29^\prime15\rlap{.}^{\prime\prime}4$ & G & 16.50 & 0.054594$^{9}$             & Y & N & N & N \\
0637$-$752&                        & $06^\mathrm {h}35^\mathrm {m}46\rlap{.}^\mathrm {s}5079$ & $-75^\circ16^\prime16\rlap{.}^{\prime\prime}814$ & Q & 15.75 & 0.653$^{10}$   & Y & N & N & N \\
1104$-$445&                        & $11^\mathrm {h}07^\mathrm {m}08\rlap{.}^\mathrm {s}6929$ & $-44^\circ49^\prime07\rlap{.}^{\prime\prime}567$ & Q & 18.20 & 1.598$^{1}$    & Y & N & N & N \\
1144$-$379&                       & $11^\mathrm {h}47^\mathrm {m}01\rlap{.}^\mathrm {s}3706$ & $-38^\circ12^\prime11\rlap{.}^{\prime\prime}022$ & Q & 16.20 & 1.048$^{11}$    & N & N & N  & Y \\
1257$-$326&                       & $13^{\rm h}00^{\rm m}42\rlap{.}^{\rm s}4259$ & $-32^\circ53^\prime12\rlap{.}^{\prime\prime}110$ & Q & 18.70 & 1.256$^{12}$                & N & N & N & N \\
1313$-$333&                       & $13^\mathrm {h}16^\mathrm {m}07\rlap{.}^\mathrm {s}9859$ & $-33^\circ38^\prime59\rlap{.}^{\prime\prime}171$ & Q & 20.00 & 1.21$^{13}$     & N & Y & C & N \\
1322$-$428&   Cen A, NGC 5128  & $13^\mathrm {h}25^\mathrm {m}27\rlap{.}^\mathrm {s}6152$ & $-43^\circ01^\prime08\rlap{.}^{\prime\prime}805$ & G & 7.84 & 0.001825$^{14}$     & Y & Y & Y  & Y \\
1323$-$526&   PMN J1326-5256   & $13^\mathrm {h}46^\mathrm {m}48\rlap{.}^\mathrm {s}70$ & $-52^\circ56^\prime22\rlap{.}^{\prime\prime}0$ & U & \dots & \ldots                 & N & Y & C$^{30}$ & N \\
1333$-$337&  IC 4296    & $13^\mathrm {h}36^\mathrm {m}39\rlap{.}^\mathrm {s}05$ & $-33^\circ57^\prime57\rlap{.}^{\prime\prime}2$ & G & 11.61 & 0.012465$^{15}$               & N & N & N & N \\
1424$-$418&             & $13^\mathrm {h}27^\mathrm {m}56\rlap{.}^\mathrm {s}2975$ & $-42^\circ06^\prime19\rlap{.}^{\prime\prime}437$ & Q & 17.7 & 1.522$^{16}$               & Y & Y & Y & N \\
1454$-$354&                 & $14^\mathrm {h}57^\mathrm {m}26\rlap{.}^\mathrm {s}7117$ & $-35^\circ39^\prime09\rlap{.}^{\prime\prime}971$ & Q & 19.50 & 1.424$^{17}$          & Y & Y & Y$^{31}$ & Y \\
1501$-$343& PMN J1505-3432  & $15^\mathrm {h}05^\mathrm {m}02\rlap{.}^\mathrm {s}4$ & $-34^\circ32^\prime57\rlap{.}^{\prime\prime}0$ & U & \ldots &  \ldots                   & N & Y & Y$^{31}$ & N \\
1549$-$790&         & $15^\mathrm {h}56^\mathrm {m}58\rlap{.}^\mathrm {s}8697$ & $-79^\circ14^\prime04\rlap{.}^{\prime\prime}281$ & G & 18.50 & 0.1501$^{18}$                 & Y & N & N & N \\
1610$-$771&               & $16^\mathrm {h}17^\mathrm {m}49\rlap{.}^\mathrm {s}2726$ & $-77^\circ17^\prime18\rlap{.}^{\prime\prime}467$ & Q & 19.00 & 1.71$^{19}$             & Y & N & N & N \\
1714$-$336&             & $17^\mathrm {h}17^\mathrm {m}36\rlap{.}^\mathrm {s}0293$ & $-33^\circ42^\prime08\rlap{.}^{\prime\prime}829$ & B &  & \ldots                         & N & Y & Y$^{31}$ & N \\
1716$-$771&             & $17^\mathrm {h}23^\mathrm {m}50\rlap{.}^\mathrm {s}51$ & $-77^\circ13^\prime50\rlap{.}^{\prime\prime}1$ & U & \ldots & \ldots                  & N & Y & C$^{32}$ & N \\
1718$-$649& NGC 6328    & $17^\mathrm {h}23^\mathrm {m}41\rlap{.}^\mathrm {s}0296$ & $-65^\circ00^\prime36\rlap{.}^{\prime\prime}615$ & G & 13.16 & 0.014428$^{20}$                & Y & N & N & N \\
1733$-$565&               & $17^\mathrm {h}37^\mathrm {m}35\rlap{.}^\mathrm {s}7706$ & $-56^\circ34^\prime03\rlap{.}^{\prime\prime}155$ & G & 18.00 & $0.098^{21}$            & Y & N & N & N \\
1759$-$396&           & $18^\mathrm {h}02^\mathrm {m}42\rlap{.}^\mathrm {s}680$ & $-39^\circ40^\prime07\rlap{.}^{\prime\prime}905$ & Q & ? & 0.296$^{22}$                     & N & Y & Y & Y* \\
1804$-$502& PMN J1808-5011& $18^\mathrm {h}08^\mathrm {m}13\rlap{.}^\mathrm {s}90$ & $-50^\circ11^\prime54\rlap{.}^{\prime\prime}0$ & Q & \ldots & \ldots                     & N & Y & C$^{32}$ & N \\
1814$-$637&                         & $18^\mathrm {h}19^\mathrm {m}35\rlap{.}^\mathrm {s}0023$ & $-63^\circ45^\prime48\rlap{.}^{\prime\prime}189$ & G & 18.00 & 0.06270$^{23}$& N & N & N & N \\
1933$-$400&              & $19^\mathrm {h}37^\mathrm {m}16\rlap{.}^\mathrm {s}2173$ & $-39^\circ58^\prime01\rlap{.}^{\prime\prime}552$ & Q & 18.00 & 0.965$^{24}$             & N & Y & Y & N \\
1954$-$388&            & $19^\mathrm {h}57^\mathrm {m}59\rlap{.}^\mathrm {s}8192$ & $-38^\circ45^\prime06\rlap{.}^{\prime\prime}356$ & Q & 17.70 & 0.63$^{25}$                & Y & N & N & N \\
2005$-$489&                       & $20^\mathrm {h}09^\mathrm {m}25\rlap{.}^\mathrm {s}3906$ & $-48^\circ49^\prime53\rlap{.}^{\prime\prime}720$ & B & 15.30 & 0.0710$^{26}$   & N & Y & Y$^{33}$ & Y \\
2027$-$308& ESO 462-G 027                          & $20^\mathrm {h}30^\mathrm {m}57\rlap{.}^\mathrm {s}934$ & $-30^\circ39^\prime24\rlap{.}^{\prime\prime}35$ & G & ? &\ldots& N & Y & Y$^{31}$ & N \\
2052$-$474&                & $20^\mathrm {h}56^\mathrm {m}16\rlap{.}^\mathrm {s}3598$ & $-47^\circ14^\prime47\rlap{.}^{\prime\prime}627$ & Q & 19.10 & 1.489$^{27}$           & Y & Y & Y & Y \\
2106$-$413&                         & $21^\mathrm {h}09^\mathrm {m}33\rlap{.}^\mathrm {s}1885$ & $-41^\circ10^\prime20\rlap{.}^{\prime\prime}605$ & Q & 21.00 & 1.058$^{16}$  & Y & N & N & N \\
2149$-$306&                         & $21^\mathrm {h}51^\mathrm {m}55\rlap{.}^\mathrm {s}5239$ & $-30^\circ27^\prime53\rlap{.}^{\prime\prime}697$ & Q & 18.40 & 2.345$^{3}$   & N & N & N & N \\
2152$-$699&  ESO 075-G 041          & $21^\mathrm {h}57^\mathrm {m}05\rlap{.}^\mathrm {s}9805$ & $-69^\circ41^\prime23\rlap{.}^{\prime\prime}685$ & G & 1430 & 0.028273$^{28}$& N & N & N & N \\
2155$-$304&                       & $21^\mathrm {h}58^\mathrm {m}52\rlap{.}^\mathrm {s}0651$ & $-30^\circ13^\prime32\rlap{.}^{\prime\prime}118$ & B & 14.00 & 0.116$^{29}$    & N & Y & Y & Y \\
2204$-$540&                  & $22^\mathrm {h}07^\mathrm {m}43\rlap{.}^\mathrm {s}7332$ & $-53^\circ46^\prime33\rlap{.}^{\prime\prime}820$ & Q & 18.00 & 1.206$^{3}$          & Y & N & N & Y \\
2326$-$477&              & $23^\mathrm {h}29^\mathrm {m}17\rlap{.}^\mathrm {s}7043$ & $-47^\circ30^\prime19\rlap{.}^{\prime\prime}115$ & Q & 16.79 & 1.2990$^{1}$             & Y & N & N & N \\
2355$-$534&               & $23^\mathrm {h}57^\mathrm {m}53\rlap{.}^\mathrm {s}2661$ & $-53^\circ11^\prime13\rlap{.}^{\prime\prime}689$ & Q & 17.80 & 1.0060$^{27}$           & N & N & N & N \\
\tableline
\enddata

\tablenotetext{a}{ IAU source designation.}
\tablenotetext{b}{ Alternative source name where appropriate.}
\tablenotetext{c}{ Right ascension and declination (J2000.0).}
\tablenotetext{d}{ The optical counterpart, denoted as follows: (G) galaxy,
(Q) quasar,  (B) BL\,Lac object, or (U) unclassified. 
}
\tablenotetext{e}{ Optical magnitude.}
\tablenotetext{f}{ Redshift.}
\tablenotetext{g}{Y= Detected, N= Not Detected, C=Candidate. Those detections and candidates without a reference indicated are from \citet{Hartman1999}.}
\tablenotetext{h}{Based on the LAT 3-month list. ``*" denotes a low confidence detection.}

\tablerefs{
(1) \citet{Peterson1976} 
(2) \citet{Wisotzki2000}
(3) \citet{Hewitt1989}
(4) \citet{Morton1978}
(5) \citet{Stickel1994}
(6) \citet{Sulentic2004}
(7) \citet{Lauberts1989}
(8) \citet{Keel1985}
(9) \citet{Quintana1995}
(10) \citet{Huntstead1978}
(11) \citet{Stickel1989}
(12) \citet{Perlman1998}
(13) \citet{Jauncey1982}
(14) \citet{Graham1978}
(15) \citet{Smith2000}
(16) \citet{White1988}
(17) \citet{Jackson2002}
(18) \citet{Tadhunter2001}
(19) \citet{Huntstead1980}
(20) HI Parkes All Sky Survey Final Catalog
(21) \citet{Tadhunter1993}
(22) \citet{Liang2003}
(23) \citet{Danziger1979b}
(24) \citet{Drinkwater1997}
(25) \citet{Browne1975}
(26) \citet{Falomo1987}
(27) \citet{Jauncey1984}
(28) \citet{daCosta1991}
(29) \citet{Falomo1993}
(30) \citet{Bignall2008}
(31) \citet{Sowards-Emmerd2004}
(32) \citet{Tornikoski2002}
(33) \citet{Lin1997}
}
\end{deluxetable}

\clearpage

\begin{deluxetable}{@{\extracolsep\fill}c@{\extracolsep\fill}c@{\extracolsep\fill}c@{\extracolsep\fill}c@{\extracolsep\fill}c@{\extracolsep\fill}c@{\extracolsep\fill}c@{\extracolsep\fill}c@{\extracolsep\fill}c@{\extracolsep\fill}c@{\extracolsep\fill}c@{\extracolsep\fill}c}
\tabletypesize{\tiny}
\tablewidth{0pt}
\tablecaption{Source Structure. \label{table:sourcestructure}}
\tablehead{
\colhead{Source} &
\colhead{Epoch} &
\colhead{Contour\tablenotemark{a}} &
\colhead{$S_{peak}$} &
\colhead{$S_{total}$} &
\colhead{$\theta_{maj}$} &
\colhead{$\theta_{min}$} &
\colhead{P.A.} &
\colhead{Structure} &
\colhead{$T_b$\tablenotemark{b,c}} &
\colhead{Core Luminosity} &
\colhead{Total Luminosity} \\
\colhead{} &
\colhead{yyyy-mm-dd} &
\colhead{($\mathrm{mJy\,beam^{-1}}$)} &
\colhead{($\mathrm{Jy\,beam^{-1}}$)}&
\colhead{($\mathrm{Jy}$)}  &
\colhead{($\mathrm{mas}$)} &
\colhead{($\mathrm{mas}$)} &
\colhead{($^\circ$)} &
\colhead{} &
\colhead{($\mathrm K$)} &
\colhead{($\mathrm{W\,m^{-2}}$)} &
\colhead{($\mathrm{W\,m^{-2}}$)} 
}
\startdata
0047$-$579 & 2007-11-10 & 1.32 & 1.04 & 1.43 & 2.27 & 0.68 & 6.3 & SS & 4.2E+11 & 2.68E+28 & 3.08E+28 \\
0208$-$512 & 2007-11-10 & 1.62 & 2.23 & 2.54 & 1.83 & 0.71 & 6.2 & SS & 6.2E+12 & 1.12E+28 & 1.26E+28 \\
0332$-$403 & 2008-02-07 & 0.30 & 0.43 & 0.46 & 4.79 & 3.62 & -64.6 & SS & 2.3E+10 & 5.67E+27 & 5.75E+27 \\
0405$-$385 & 2008-02-07 & 2.26 & 1.29 & 1.35 & 3.98 & 2.68 & -69.1 & SS & 1.6E+11 & 1.24E+28 & 1.26E+28 \\
0438$-$436 & 2008-02-07 & 1.47 & 0.83 & 1.42 & 4.58 & 3.21 & -70.1 & SS & 7.0E+10 & 6.02E+28 & 9.59E+28 \\
0454$-$463 & 2007-11-10 & 4.20 & 2.52 & 3.65 & 2.12 & 0.59 & 10.0 & SS & 3.0E+12 & 8.77E+27 & 1.22E+28 \\
0506$-$612 & 2007-11-10 & 0.96 & 0.85 & 0.93 & 1.62 & 0.74 & 7.1 & SS & 1.8E+12 & 5.52E+27 & 5.79E+27 \\
0518$-$458 & 2007-11-10 & 1.26 & 0.50 & 0.87 & 2.24 & 0.74 & 4.9 & SS & 2.1E+11 & 1.56E+24 & 2.28E+24 \\
0521$-$365 & 2007-11-10 & 1.50 & 1.07 & 1.62 & 2.54 & 0.69 & 2.6 & SS & 1.2E+11 & 4.04E+24 & 1.09E+25 \\
0537$-$441 & 2007-11-10 & 3.30 & 4.89 & 5.13 & 2.17 & 0.70 & 11.6 & SS & 1.1E+14 & 1.84E+28 & 1.93E+28 \\
0625$-$354 & 2007-11-10 & 0.30 & 0.29 & 0.36 & 2.91 & 0.90 & 4.7 & SS & 8.1E+10 & 1.99E+24 & 2.35E+24 \\
0637$-$752 & 2008-02-07 & 1.86 & 2.67 & 3.47 & 3.95 & 0.93 & 66.5 & SS & 1.9E+11 & 5.30E+27 & 6.00E+27 \\
1104$-$445 & 2007-11-10 & 2.10 & 1.06 & 1.42 & 3.09 & 0.79 & 1.6 & SS & 8.4E+11 & 1.88E+28 & 2.28E+28 \\
1144$-$379 & 2008-02-07 & 1.30 & 1.68 & 1.77 & 5.14 & 3.24 & -75.1 & SS & 8.3E+11 & 9.50E+27 & 9.92E+27 \\
1257$-$326 & 2008-02-07 & 0.27 & 0.13 & 0.18 & 3.70 & 0.78 & -0.5 & SS & 8.1E+10 & 1.21E+27 & 1.59E+27 \\
1313$-$333 & 2008-02-07 & 0.57 & 0.68 & 0.77 & 3.21 & 0.59 & -0.2 & SS & 3.6E+12** & 5.49E+27 & 6.18E+27 \\
1322$-$428 & 2007-11-10 & 1.20 & 0.60 & 2.75 & 1.64 & 0.41 & 7.9 & DS & 3.7E+10 & 8.03E+21 & 2.29E+22 \\
1323$-$526 & 2007-11-10 & 0.81 & 1.00 & 1.08 & 2.16 & 0.83 & 8.4 & SS & 1.3E+12* & \ldots & \ldots \\
1333$-$337 & 2008-02-07 & 0.36 & 0.17 & 0.24 & 3.74 & 0.97 & -7.4 & DS & 4.8E+10 & 6.07E+22 & 7.67E+22 \\
1424$-$418 & 2007-11-10 & 0.84 & 1.15 & 1.58 & 2.38 & 0.68 & 3.4 & SS & 3.0E+12 & 1.93E+28 & 2.25E+28 \\
1454$-$354 & 2007-11-10 & 0.27 & 0.37 & 0.44 & 2.87 & 0.75 & 3.6 & SS & 3.3E+11 & 4.86E+27 & 5.30E+27 \\
1501$-$343 & 2007-11-10 & 0.30 & 0.24 & 0.24 & 2.76 & 0.68 & 3.6 & C & 1.9E+11*** & \ldots & \ldots \\
1549$-$790 & 2008-02-07 & 0.48 & 0.28 & 1.37 & 1.96 & 0.62 & -0.3 & DS & 1.5E+10 & 2.67E+25 & 9.76E+25 \\
1610$-$771 & 2008-02-07 & 2.19 & 0.86 & 1.83 & 1.40 & 0.50 & 5.2 & SS & 3.5E+11 & 2.44E+28 & 3.48E+28 \\
1714$-$336 & 2007-11-10 & 0.51 & 0.61 & 1.01 & 4.46 & 3.46 & -72.6 & SS & 1.8E+09* & \ldots & \ldots \\
1716$-$771 & 2007-11-10 & 0.57 & 0.12 & 0.14 & 1.33 & 0.83 & 13.5 & SS & 6.6E+10 & 5.39E+22 & 6.02E+22 \\
1718$-$649 & 2008-02-07 & 3.90 & 1.07 & 3.36 & 2.13 & 0.45 & -12.4 & Irr & 1.1E+11* & \ldots & \ldots \\
1733$-$565 & 2008-02-07 & 0.18 & 0.15 & 0.18 & 2.63 & 0.79 & -4.0 & DS & 4.5E+10 & 3.75E+24 & 4.04E+24 \\
1759$-$396 & 2007-11-10 & 1.00 & 1.27 & 1.42 & 3.00 & 0.67 & 5.7 & SS & 2.4E+12 & 3.35E+26 & 3.70E+26 \\
1804$-$502 & 2007-11-10 & 0.39 & 0.32 & 0.33 & 2.60 & 0.67 & 8.6 & SS & 2.4E+12** & 5.23E+27 & 5.37E+27 \\
1814$-$637 & 2008-02-07 & 1.70 & 0.37 & 0.93 & 1.92 & 0.61 & 2.3 & DS & 1.7E+11 & 3.73E+24 & 8.12E+24 \\
1933$-$400 & 2007-11-10 & 1.50 & 1.15 & 1.30 & 3.30 & 0.90 & 23.3 & SS & 2.0E+12 & 5.32E+27 & 5.93E+27 \\
1954$-$388 & 2008-02-07 & 1.92 & 1.28 & 2.30 & 3.09 & 0.38 & 5.0 & SS & 1.5E+12 & 1.66E+27 & 3.64E+27 \\
2005$-$489 & 2007-11-10 & 0.72 & 0.44 & 0.65 & 3.66 & 1.09 & 36.6 & SS & 2.5E+10 & 6.13E+24 & 7.36E+24 \\
2027$-$308 & 2008-06-09 & 0.75 & 0.09 & 0.12 & 3.43 & 1.39 & -6.9 & SS & 4.1E+10*** & \ldots & \ldots \\
2052$-$474 & 2008-02-07 & 0.60 & 1.49 & 1.67 & 3.19 & 0.47 & 12.8 & SS & 2.0E+12 & 2.21E+28 & 2.25E+28 \\
2106$-$413 & 2008-02-07 & 1.50 & 0.78 & 1.30 & 2.51 & 0.46 & 10.4 & SS & 1.8E+11 & 6.69E+27 & 7.46E+27 \\
2149$-$306 & 2008-03-28 & 2.51 & 0.90 & 1.34 & 3.31 & 0.60 & -0.9 & SS & 9.3E+11 & 4.06E+28 & 5.56E+28 \\
2152$-$699 & 2008-02-07 & 0.33 & 0.38 & 0.51 & 3.38 & 0.75 & -4.0 & SS & 4.7E+10 & 7.59E+23 & 8.60E+23 \\
2155$-$304 & 2008-03-28 & 0.60 & 0.39 & 0.51 & 3.98 & 0.62 & -1.5 & SS & 3.3E+10 & 1.49E+25 & 1.64E+25 \\
2204$-$540 & 2008-02-07 & 1.95 & 0.84 & 1.09 & 2.12 & 0.51 & -2.4 & SS & 6.0E+11 & 7.09E+27 & 8.67E+27 \\
2326$-$477 & 2007-11-10 & 2.52 & 0.52 & 1.06 & 2.01 & 0.60 & 6.8 & SS & 3.4E+11 & 5.61E+27 & 1.02E+28 \\
2355$-$534 & 2008-02-07 & 0.84 & 1.44 & 1.68 & 4.08 & 0.88 & -11.3 & SS & 1.2E+12 & 7.50E+27 & 8.50E+27 \\
\tableline
\enddata
\tablenotetext{a}{Usually $3\times$RMS noise in image}
\tablenotetext{b}{``*" indicates that $z=0$ was used as a limit.}
\tablenotetext{c}{``**'' indicates that the value has been calculated from theoretical SNR limits as
described in section 6.4. These limits were checked using all possible
combinations of major axis and axial ratio using the difwrap package
(Lovell 1998) and were found to be consistent.}
\end{deluxetable}

\clearpage

\begin{deluxetable}{@{\extracolsep\fill}c@{\extracolsep\fill}c@{\extracolsep\fill}c@{\extracolsep\fill}c@{\extracolsep\fill}c@{\extracolsep\fill}c@{\extracolsep\fill}c@{\extracolsep\fill}c@{\extracolsep\fill}c@{\extracolsep\fill}}
\tabletypesize{\tiny}
\tablewidth{0pt}
\tablecaption{Source Structure of Tapered Images. \label{table:sourcestructuretapered}}
\tablehead{
\colhead{Source} &
\colhead{Epoch} &
\colhead{Contour} &
\colhead{$S_{peak}$}&
\colhead{$S_{total}$} &
\colhead{$\theta_{maj}$} &
\colhead{$\theta_{min}$} &
\colhead{P.A.} &
\colhead{Taper\tablenotemark{a}}\\
\colhead{} &
\colhead{yyyy-mm-dd} &
\colhead{($\mathrm{mJy\,beam^{-1}}$)} &
\colhead{($\mathrm{Jy\,beam^{-1}}$)}&
\colhead{($\mathrm{Jy}$)}  &
\colhead{($\mathrm{mas}$)} &
\colhead{($\mathrm{mas}$)} &
\colhead{($^\circ$)} &
\colhead{}
}
\startdata
0047$-$579 & 2007-11-10 & 1.25 & 1.25 & 1.42 & 4.32 & 3.71 &  86.2 & 100 \\
0208$-$512 & 2007-11-10 & 1.15 & 2.32 & 2.51 & 4.32 & 3.52 & -82.8 & 100 \\
0454$-$463 & 2007-11-10 & 2.49 & 3.11 & 3.52 & 2.78 & 1.22 &  11.2 & 100 \\
0506$-$612 & 2007-11-10 & 0.90 & 0.90 & 0.97 & 5.61 & 4.70 & -83.8 & 50 \\
0625$-$35  & 2007-11-10 & 3.00 & 0.32 & 0.36 & 5.59 & 4.59 & -71.0 & 50\\
0637$-$752 & 2008-02-07 & 1.50 & 3.12 & 3.51 & 6.55 & 5.24 & -47.3 & 50 \\
1144$-$379 & 2008-02-07 & 0.55 & 1.69 & 1.77 & 7.33 & 4.45 & -71.8 & 50\\
1424$-$418 & 2007-11-10 & 1.37 & 1.37 & 1.53 & 4.79 & 3.64 & -60.0 & 75\\
1454$-$354 & 2007-11-10 & 0.30 & 0.40 & 0.43 & 4.89 & 3.69 & -60.4 & 100\\
1549$-$790 & 2008-02-07 & 0.63 & 0.27 & 1.35 & 5.09 & 3.67 & -79.5 & 100 \\
1610$-$771 & 2008-02-07 & 4.20 & 1.40 & 1.79 & 5.31 & 4.14 &  65.0 & 100\\
1814$-$637 & 2008-02-07 & 1.70 & 0.48 & 0.93 & 7.16 & 6.24 &  62.4 & 35\\
2005$-$489 & 2007-11-10 & 0.53 & 0.53 & 0.64 & 7.56 & 5.09 &  84.6 & 35\\
\tableline
\enddata
\tablenotetext{a}{Baseline length in $M\lambda$ at which the visibility data were downweighted to 10\,\%.}
\end{deluxetable}

\clearpage

\begin{deluxetable}{lrr}
\tabletypesize{\footnotesize}
\tablecaption{Distribution of Morphology\label{table:sourcedistribution}}
\tablewidth{0pt}
\tablehead{
\colhead{Structure\tablenotemark{a} } & 
\colhead{Number} & 
\colhead{\%} 
}
\startdata
SS  & 36 & 84 \\
C   &  1 &  2 \\
DS  &  5 & 12 \\
Irr &  1 & 2 \\
\enddata

\tablenotetext{a}{Classification of the structure, as follows: (C) compact,
(SS) single-sided, 
(DS) double-sided, 
or (Irr) irregular.}
\end{deluxetable}

\clearpage

\begin{deluxetable}{rrrrrrc}
\tablecaption{Opening Angles of TANAMI sources\label{table:openingangle}}
\tablewidth{0pt}
\tablehead{
\colhead{Source} & \colhead{Class} & \colhead{r\tablenotemark{1}} & \colhead{$\theta\tablenotemark{2}$} & \colhead{Component Size} & \colhead{Opening Angle} & \colhead{Limit?\tablenotemark{3}} \\
\colhead{} & \colhead{} & \colhead{(mas)}  & \colhead{(degrees)} & \colhead{(mas)} & \colhead{(degrees)} & \colhead{}
}
\startdata
0047$-$579 & Q & 2.11 & $-$13.0 & 0.26 & 7.2 & Y \\
0208$-$512 & B & 2.93 & 24.3 & 1.88 & 35.6 & N\\
0438$-$436 & Q & 6.54 & $-$44.8 & 1.36 & 11.9 & N \\
0506$-$612 & Q & 3.02 & $-$46.1 & 0.19 & 3.7 & Y \\
0521$-$365 & B & 3.04 & $-$37.2 & 0.36 & 6.9 & N \\
0537$-$441 & Q & 2.57 & 57.2 & 1.41 & 30.9 & N \\
0637$-$752 & Q & 6.08 & 1.5 & 0.42 & 4.0 & Y \\
1104$-$445 & Q & 3.12 & 27.2 & 2.14 & 38.2 & N \\
1144$-$379 & Q & 1.02 & $-$54.3 & 0.65 & 36.4 & N \\
1257$-$326 & Q & 2.8 & $-$36.4 & 1.51 & 30.7 & N \\
1313$-$333 & Q & 3.74 & $-$7.5 & 2.45 & 36.3 & N \\
1323$-$526 & U & 3.26 & 83.2 & 0.41 & 7.3 & N \\
1424$-$418 & Q & 0.98 & 42.5 & 0.87 & 47.9 & N \\
1454$-$354 & Q & 1.77 & 2.6 & 0.87 & 27.5 & N \\
1610$-$771 & Q & 2.16 & $-$48.1 & 0.40 & 10.7 & N \\
1714$-$336 & B & 13.61 & 32.7 & 4.73 & 19.7 & N \\
1716$-$771 & U & 2.3 & 49.8 & 0.30 & 7.4 & Y \\
1759$-$396 & Q & 2.61 & $-$74.6 & 1.67 & 36.2 & N \\
1804$-$502 & Q & 2.29 & 54.1 & 0.96 &23.9 & N \\
1933$-$400 & Q & 2.39 & $-$51.0 & 0.31 & 7.6 & Y \\
1954$-$388 & Q & 1.55 & $-$44.1 & 0.45 & 16.5 & N \\
2005$-$489 & B & 1.52 & 31.8 & 1.18 & 42.4 & N \\
2052$-$474 & Q & 1.91 & $-$31.0 & 0.27 & 8.0 & N \\
2106$-$413 & Q & 3.1 & 1.7 & 0.47 & 8.6 & Y \\
2149$-$306 & Q & 2.49 & $-$28.1 & 0.26 & 6.0 & Y \\
2155$-$304 & B & 2.72 & $-$56.2 & 1.74 & 35.4 & N \\
2204$-$540 & Q & 2.14 & 64.6 & 0.30 & 8.0 & N \\
2326$-$477 & Q & 3.47 & $-$11.0 & 0.30 & 5.0 & Y \\
2355$-$534 & Q & 3.88 & 37.9 & 0.49 & 7.3 & N \\
\enddata
\tablenotetext{1}{Radial distance between core and fitted component}
\tablenotetext{2}{Angle between core and fitted component}
\tablenotetext{3}{Flag indicating whether the innermost jet component is unresolved, resulting in an upper limit on the opening angle. Y= yes, N= no}
\end{deluxetable}

\end{document}